\documentclass[preprint,prb,showpacs,superscriptaddress]{revtex4-1}
\pdfoutput=1
\usepackage{amsmath}
\usepackage{graphicx}
\usepackage{amssymb}
\usepackage{times}
\usepackage{color}
\definecolor{lr}{rgb}{1.0,0.3,0.3}
\definecolor{dg}{rgb}{0.0,0.5,0.0}

\makeatletter

\@ifundefined{textcolor}{}
{%
 \definecolor{BLACK}{gray}{0}
 \definecolor{WHITE}{gray}{1}
 \definecolor{RED}{rgb}{1,0,0}
 \definecolor{GREEN}{rgb}{0,1,0}
 \definecolor{BLUE}{rgb}{0,0,1}
 \definecolor{CYAN}{cmyk}{1,0,0,0}
 \definecolor{MAGENTA}{cmyk}{0,1,0,0}
 \definecolor{YELLOW}{cmyk}{0,0,1,0}
 }

\makeatother

\begin{document}

\title{Theoretical unification of hybrid-DFT and DFT+$U$ methods for the treatment of localized orbitals}

\author{Viktor Iv\'ady}
\email{vikiv@ifm.liu.se} 
\affiliation{Department of Physics, Chemistry and Biology, Link\"oping
  University, SE-581 83 Link\"oping, Sweden}
\affiliation{Wigner Research Centre for Physics, Hungarian Academy of Sciences,
  PO Box 49, H-1525, Budapest, Hungary}

\author{Rickard Armiento}
\affiliation{Department of Physics, Chemistry and Biology, Link\"oping
  University, SE-581 83 Link\"oping, Sweden}
  
\author{Kriszti\'an Sz\'asz} 
\affiliation{Wigner Research Centre for Physics, Hungarian Academy of Sciences,
  PO Box 49, H-1525, Budapest, Hungary}

\author{Erik Janz\'en}
\affiliation{Department of Physics, Chemistry and Biology, Link\"oping
  University, SE-581 83 Link\"oping, Sweden}

\author{Adam Gali} 
\affiliation{Wigner Research Centre for Physics, Hungarian Academy of Sciences,
  PO Box 49, H-1525, Budapest, Hungary}
\affiliation{Department of Atomic Physics, Budapest University of
  Technology and Economics, Budafoki \'ut 8., H-1111 Budapest,
  Hungary}

\author{Igor A. Abrikosov}
\affiliation{Department of Physics, Chemistry and Biology, Link\"oping
  University, SE-581 83 Link\"oping, Sweden}

\pacs{61.72.J-, 61.82.Fk, 71.15.Mb, 76.30.-v}

\begin{abstract}
We formulate the on-site occupation dependent exchange correlation energy and effective potential of hybrid functionals for localized states and connect them to the on-site correction term of the DFT+$U$ method. Our derivation provides a theoretical justification for adding a DFT+$U$-like onsite potential in hybrid DFT calculations to resolve issues caused by overscreening of localized states. The resulting scheme, hybrid-DFT+V$_{\textup{w}}$, is tested for chromium impurity in wurtzite AlN and vanadium impurity in 4H-SiC, which are paradigm examples of systems with different degree of localization between host and impurity orbitals.

\end{abstract}
\maketitle

\section{Introduction}

To investigate solid state systems based on first principles quantum mechanical simulations has been a rapidly developing field of physics for many decades thanks to the increasing amount of available computational resources and a significant improvement in the description of systems of many interacting electrons.  One of the largest families of first principles techniques is the density functional theory (DFT), formulated by P.~Hohenberg, W.~Kohn and L.~J.~Sham in 1964 \citep{Hohenberg64,Kohn65} . Even using local or semilocal approximations for the unknown exchange-correlation energy, e.g.\ the local density approximation (LDA) \citep{Kohn65} or the generalized gradient approximation \citep{PW91,PBE} (GGA), this theory can predict physical observables with reasonable accuracy and relatively low computational cost for a large set of systems \citep{Kohn_Nobel}. However, in spite of their great success, this type of approximations suffers from a few long standing closely related problems; the self-interaction error\citep{Perdew81}, the absence of derivative discontinuity in the exchange correlation potential at integer occupation numbers \citep{Perdew83,Sham83,Perdew91} and qualitative errors appearing for highly correlated systems\citep{Anisimov91,Anisimov93,LDAUrev}. Since methods based on higher level of theory, for instance the GW approximation\citep{Hedin} in many-body perturbation theory (MBPT) and dynamic mean field theory\citep{Georges92,DMFTrev96,DMFTrev} (DMFT), are computationally demanding, efforts for improving DFT based techniques are still highly needed and potentially of great impact.

One way of improving the description over local and semilocal approximations in DFT is the use of hybrid functionals \citep{Kummel08}, which mix the exact exchange energy of the Kohn-Sham (KS) particles with the (semi)local approximation of exchange energy of DFT. The concept of hybrid functionals was derived from the adiabatic connection formalism by Becke \citep{Becke93a} in 1993. The hybrid formalism makes it possible to improve the exchange-correlation energy by mixing the non-local exact exchange energy of the Kohn-Sham orbitals and the (semi)local exchange energy functional in the theoretical framework of the generalized Kohn-Sham scheme \citep{LevyGKS}. The ratio of the exact exchange energy part can be related to the order of the perturbation theory needed to describe the chosen system properly\citep{PerdewMix}. For materials with \emph{sp} hybridized orbitals this ratio, i.e., the mixing parameter of the hybrid functional, is approximately $ 0.25 $.

Since the birth of the hybrid functional scheme, semi-empirical functional forms with different number of fitting parameters were proposed and adjusted to describe large sets of molecules \citep{Becke93b,B3LYP,HSE03}. The B3LYP functional have become a successful tool in the field of quantum chemistry and as a result is now in frequent use. The use of  hybrid functionals in the solid state community has been partially hindered by technical difficulties, which originate from the treatment of the long ranged and non-local exact exchange potential for periodic solids \citep{Kummel08}. The introduction of range separated hybrid functionals, i.e., HSE06 \citep{HSE03,HSE06}, made it possible to overcome these difficulties. By now, HSE06 has become a state-of-art tool in the field of solid state physics. The success of hybrids for solid state applications can be understood as a consequence of the reduced self-interaction error and the introduction of the derivative discontinuity of the exchange correlation functional. Over the last few years, the remarkable predictive power of the non-empirical optimally tuned hybrid functionals has drawn the attention of the scientific community\citep{Baer10,Srebro12,KronikJCTC12, KronikPRL12,KronikPRB13,KummelJCP13,Atalla13,Richter13}. In such approaches the features of the exact functional are enforced in the case of the approximate density functionals, which has turned out to be a generally successful way to improve the first principles description \citep{Cococcioni05,Cococcioni06,LanyZunger09,LanyZunger10,Dabo,Baer10,Andrade11,Xiao11,Gaiduk12,Baer10,Srebro12,KronikJCTC12, KronikPRL12,KronikPRB13,KummelJCP13,Atalla13,Richter13, Kronik13}. On the other hand, one of the drawbacks with hybrid functionals is that the homogeneous and global mixing of the two kinds of exchange terms is governed by a single mixing parameter. Perdew \emph{et al.}~\citep{Perdew07} pointed out that such behavior hinders the correct description of space dependent phenomena. To overcome this shortcoming the so called local hybrids were suggested \citep{Jaramillo03,Arbuznikov07,Bahmann07,Kaupp07,Janesko08,Krukau08,Arbuznikov09}. 

The treatment of strongly interacting and correlated particles is especially problematic in (semi)local-DFT. To reproduce band structure closer to experiment a common remedy has been to apply the DFT+$U$ scheme \cite{LDAUrev, Anisimov91_late, Anisimov93, Liechtenstein95, Dudarev98, Cococcioni05, RinkeLDAU10}. In this method the treatment of the subset of correlated orbitals has a direct connection to the advanced GW approximation of MBPT \citep{LDAUrev,RinkeLDAU10}. On the other hand, large part of the exchange and correlation effects are still described in (semi)local-DFT, which suffers from the self-interaction error and the absence of the derivative discontinuity. In the case of correlated points defect in the host of conventional semiconductors, neither hybrid functionals nor DFT+$U$ can provide an accurate description, however, a corrected hybrid functional, presented in a previous paper, the HSE06+V$_{\textup{w}}$, can overcome the difficulties \citep{Ivady13}. In this article we present the theoretical motivation and foundation of this method, as well as deeper insights into the connection between the hybrid-DFT and DFT+$U$ methods. The proposed hybrid-DFT+V$_{\textup{w}}$ scheme provides an alternative solution to the problems arise from the homogeneous mixing used in hybrid functionals.

The article is organized as follows: Section II summarizes the foundations of the DFT+U method and hybrid functionals. In Section III we establish a connection between these two methods for localized orbitals and discuss the consequences. The following topics are presented in the subsections: the effect of hybrid functionals on localized orbitals, an introduction to hybrid-DFT+V$_\textup{w}$, self-consistent determination of parameter $w$ and finally we discuss the band gap in DFT+$U$ and hybrid-DFT schemes. In Section IV the use of hybrid-DFT+V$_{\textup{w}}$ and its effects on localized orbitals are presented in the case of substitutional chromium at aluminum site in w-AlN and substitutional vanadium at silicon site in 4H-SiC. In Section V we summarize our findings. 

\section{Background}

In the following we give a brief summary of the DFT+$U$ scheme and hybrid functionals.

\subsection{DFT+$U$ method}

The DFT+$U$ method was introduced by Anisimov and co-workers to remedy issues in (semi)local DFT with the description of localized states, which is especially important for strongly correlated materials \citep{Anisimov91_late,Anisimov93,Liechtenstein95,Dudarev98,LDAUrev}. We now summarize this scheme, closely following the presentation by Cococcinio \emph{et~al}.\ in Ref.~\onlinecite{Cococcioni05}.

In the DFT+$U$ scheme, the DFT energy functional is extended by an on-site Hubbard-like term,
\begin{equation} \label{eq:LDAU_1}
E_{\textup{DFT+U}} \! \! \left [  \varrho\!\left ( \mathbf{r} \right )  \right ] = E_{\textup{DFT}}\! \left [  \varrho\!\left ( \mathbf{r} \right ) \right ] + E_{\textup{Hub}}^{I} \! \! \left [   \left \{  n_{mm'}^{I \sigma }  \right \} \right ]  -  E_{\textup{DC}}^{I} \! \! \left [   \left \{  n_{mm'}^{I\sigma }  \right \} \right ] ,
\end{equation}
where the energy term  $E_{\textup{DFT}}$ is the DFT total energy of the electron system,  $ E_{\textup{Hub}}^{I}$ is the Hubbard interaction energy of the localized correlated orbitals of atom $I$ and $E_{\textup{DC}}^{I} $ is the approximated DFT interaction energy of the orbitals, which must be subtracted to avoid double counting of the interaction of the corrected orbitals. To simplify notation we consider systems with one correlated atomic site, and therefore we leave out the superscript $I$. The last two terms on the right hand side depend on the on-site occupation matrix $n_{m{m}'}^{\sigma}$ defined as \citep{Cococcioni05}
\begin{equation}
n_{m{m}'}^{\sigma} = \sum_{n\mathbf{k}} f_{n\mathbf{k}}^{\sigma} \left \langle  \psi_{n \mathbf{k}}^{\sigma} \left | P_{m{m}'} \right |  \psi_{n \mathbf{k}}^{\sigma} \right \rangle,
\end{equation}
where $\psi_{n \mathbf{k}}^{\sigma}$ are the Kohn-Sham orbitals of the Kohn-Sham particles, $f_{n\mathbf{k}}^{\sigma} $ are the corresponding occupation numbers of the orbitals and $P_{m{m}'}$ are projector operators built up from localized orbitals $ \phi_m $ as
\begin{equation}
P_{m{m}'} = \left |  \phi_m  \left \rangle  \right \langle \phi_{{m}'} \right |.
\end{equation}
With the definition of the product $ C_{m;n\textbf{k}}^{\sigma } = \left \langle \phi_{m}  | \psi_{n\textbf{k}}^\sigma \right \rangle$, the on-site occupation matrix can be written as
\begin{equation}  \label{eq:LDAU_5}
  n_{m{m}'}^{\sigma } = \sum_{n\mathbf{k} } f_{n\textbf{k}}^\sigma C_{m;n\textbf{k}}^{\sigma \ast  } C_{{m}';n\textbf{k}}^{\sigma }.
\end{equation}
The atomic Hartree-Fock interaction energy can be expressed in terms of the occupation matrix elements
\begin{align} \label{eq:LDAU_2}
E_{\textup{Hub}} \! \left [ \left \{ n_{m}^{\sigma }  \right \} \right ] =& \frac{1}{2} \sum_{ \left \{ m \right \}, \sigma } \left \{   \left \langle m m_{1} \left | v_{ee} \right | {m}' m_{2} \right \rangle n_{m{m}'}^{\sigma } n_{m_{1} m_{2}}^{-\sigma }    \nonumber  \right.   \\   
  + & \left. \left (  \left \langle m m_{1} \left | v_{ee} \right | {m}' m_{2} \right \rangle   \nonumber \right. \right. \\
  - & \left. \left.  \left \langle m m_{1} \left | v_{ee} \right | m_{2} {m}' \right \rangle \right )  n_{m{m}'}^{\sigma }  n_{m_{1} m_{2}}^{\sigma }   \right \},
\end{align}
where $v_{ee}$ is the Coulomb interaction potential, 
\begin{equation} \label{eq:LDAU_9}
  v_{ee} \! \left ( \textbf{r} - {\textbf{r}}' \right ) = \frac{e^{2}}{4\pi \varepsilon_{0}} \frac{1}{\left | \textbf{r} - {\textbf{r}}' \right |} .
\end{equation}
The matrix elements of this kernel can be written as linear combinations of Slater integrals $ F^{k} $ 
\begin{equation}
\left \langle m m_{1} \left | v_{ee}  \right | {m}' m_{2} \right \rangle = \sum_{k=0}^{2l} a_{k} \! \left ( m, {m}', m_{1}, m_{2}  \right ) F^{k}.
\end{equation}
The angular integrals $ a_{k} $ are
\begin{align}
a_{k} \! \left ( m, {m}', m_{1}, m_{2} \right ) = & \frac{4\pi }{2k+1} \sum_{q=-k}^{k} \left \langle lm \left |  Y_{kq} \right | l{m}' \right \rangle    \\
 \times & \left \langle lm_{1} \left | Y_{kq}^{\ast }  \right |  lm_{2} \right \rangle ,
\end{align}
where $Y_{kq}$ are spherical harmonics. For \emph{d} electrons there are three non-vanishing integrals $ F^{0} $, $ F^{2} $ and $ F^{4} $ that can be expressed with only two parameters,
\begin{align}  \label{eq:LDAU_6}
F^{0}  = & \frac{1}{ {\left ( 2l+1 \right )}^{2} } \sum_{m,{m}'}  F^{0}_{m{m}'}  \nonumber \\
=& \frac{1}{ {\left ( 2l+1 \right )}^{2} } \sum_{m,{m}'}  \left \langle m {m}' \left | v_{ee}  \right | m {m}' \right \rangle  ,
\end{align}
\begin{equation}  \label{eq:LDAU_7}
J^{0} = \frac{1}{ 2l \left ( 2l+1 \right ) }  \sum_{m\neq {m}',{m}'}  \left \langle m {m}' \left | v_{ee}  \right | {m}' m \right \rangle = \frac{F^{2} + F^{4}}{14}  ,
\end{equation}
while the ratio of $F^{2}$ and $F^{4}$ is fixed
\begin{equation}
\frac{F^{4}}{F^{2}} \approx 0.625 .
\end{equation}
In practice, to take into account the screening effect of the other electrons in the system, these integrals are not calculated explicitly, but rather treated as parameters. The Hubbard $U$ and the Stoner $J$ parameters are the corresponding screened value of $F^{0}$  and $J^{0}$, respectively.  

The double counting term in Eq.~(\ref{eq:LDAU_1}) is a somewhat arbitrary part of the derivation of the DFT+$U$ method. There are several proposals for this term, however, in this paper, we apply the one originally used by Anisimov and coworkers\cite{Liechtenstein95}, which is the most frequently used one (see Ref.~\onlinecite{RinkeLDAU10} and references therein for more details). The total energy of the correlated orbitals in the fully localized limit (FLL)\citep{Czyzyk94,Liechtenstein95} can be obtained from Eq.~(\ref{eq:LDAU_2}) by neglecting orbital polarization effects.  It becomes
\begin{equation} \label{eq:LDAU_3}
E_{\textup{DC}} \! \left [ \left \{ n_{m}^{\sigma }  \right \} \right ] = \frac{U}{2} n \left ( n - 1 \right ) - \frac{J}{2}  \sum_{\sigma } n^{\sigma } \left ( n^{\sigma } - 1 \right ) ,
\end{equation}
where $n=n^{\uparrow}+n^{\downarrow}$ and $ n^{\sigma } = \textup{Tr}\left ( n_{m{m}'}^{\sigma } \right )$. The derivative of the total energy function Eq.~(\ref{eq:LDAU_2}) and Eq.~(\ref{eq:LDAU_3}) with respect to the occupation matrix element $n_{m{m}'}^{\sigma}$ gives the on-site correction potential to the (semi)local Kohn-Sham potential in the DFT+$U$ method,
\begin{align} \label{eq:LDAU_4}
\Delta \! V_{m{m}'}^{\sigma} = & \sum_{ \left \{ m \right \}, \sigma } \left \{   \left \langle m m_{1} \left | v_{ee} \right | {m}' m_{2} \right \rangle n_{m_{1} m_{2}}^{-\sigma }    \nonumber  \right.   \\   
  + & \left. \left (  \left \langle m m_{1} \left | v_{ee} \right | {m}' m_{2} \right \rangle   \nonumber \right. \right. \\
  - & \left. \left.  \left \langle m m_{1} \left | v_{ee} \right | m_{2} {m}' \right \rangle \right ) n_{m_{1} m_{2}}^{\sigma }   \right \}  \nonumber \\
  - &  U\left ( n - \frac{1}{2} \right ) + J \left ( n^{\sigma}  - \frac{1}{2}\right )
\end{align}

In the version of the scheme by Dudarev \emph{et al}.\citep{Dudarev98}\ the potential can be written in a more transparent form by using spherically averaged $U$ and $J$ parameters,  i.e., $ \left \langle m m_{1} \left | v_{ee} \right | {m}' m_{2} \right \rangle \approx U $ and $  \left \langle m m_{1} \left | v_{ee} \right | m_{2} {m}' \right \rangle \approx J $.  The rotationally invariant form of the total energy functional of Eq.~(\ref{eq:LDAU_1}) then becomes
\begin{align}
E_{\textup{DFT+U}} \! \! \left [  \varrho\!\left ( \mathbf{r} \right )  \right ] = & E_{\textup{DFT}}\! \left [  \varrho\!\left ( \mathbf{r} \right ) \right ]  \nonumber \\ 
+ & \frac{ U_{\textup{eff}}}{2} \left (  \sum_{m\sigma} n_{mm}^{\sigma} - \sum_{m{m}' \sigma}  n_{m{m}'}^{\sigma} n_{{m}'m}^{\sigma} \right ) ,
\end{align}
where $ \ U_{\textup{eff}} = U - J $. This equation can be further simplified by choosing the atomic basis set $ \left | \phi_{m} \right\rangle $ in such a way that the on-site occupation matrix becomes diagonal,  
\begin{equation}  \label{eq:LDAU_10}
E_{\textup{DFT+U}} \! \! \left [  \varrho\!\left ( \mathbf{r} \right )  \right ] =  E_{\textup{DFT}}\! \left [  \varrho\!\left ( \mathbf{r} \right ) \right ] + \frac{ U_{\textup{eff}}}{2}  \sum_{m\sigma} \left (  n_{m}^{\sigma} -  { \left (   n_{m}^{\sigma}   \right )}^2  \right ) ,
\end{equation}
where $n_{m}^{\sigma} = n^{\sigma}_{mm} $. From this form one can get a physically understandable and transparent  potential correction expression,
\begin{equation}  \label{eq:LDAU_8}
 \Delta \! V_{m}^{\textup{DFT+U}, \sigma}  =  U_{\textup{eff}} \left (  \frac{1}{2}  - n_{m}^{\sigma }  \right ) ,
\end{equation} 
As can be understood from this result, the major effect of the introduced Hubbard interaction term on the Kohn-Sham energies of the occupied and unoccupied correlated orbitals is to decrease and increase them by $ U_{\textup{eff}}/2 $, respectively. Thus, the so called Hubbard gap is generated between the occupied an unoccupied states. 

\subsection{ Hybrid functionals} \label{subsec:hybrid}
 
In the subsequent section we discuss two hybrid functionals, PBE0, by Adamo \emph{et al}.\citep{PBE0}, and the range separated version of this functional, the HSE06, by Heyd \emph{et al}.~\citep{HSE03,HSE06}. These functionals are widespread in solid state applications. In this subsection we give a short overview of the formulation of these functionals.

The PBE0 exchange and correlation energy functional is defined in the form
\begin{equation}
E^{\textup{PBE0}}_{\textup{xc}}  \! \left [ \varrho,\left \{ \psi_{n\textbf{k}}^\sigma   \right \} \right ] =  E_{\textup{xc}}^{\textup{PBE}} \!  \left [  \varrho \right ]  + \alpha  E_{\textup{x}}^{\textup{ex}} \! \left [ \left \{ \psi_{n\textbf{k}}^\sigma   \right \}\right ] - \alpha E_{\textup{x}}^{\textup{PBE}} \!  \left [  \varrho \right ] ,
\end{equation}
where $  \alpha  $ is the mixing parameter, $ E_{\textup{xc}}^{\textup{DFT}} \!  \left [  \varrho \! \left ( \mathbf{r} \right ) \right ] $ is the PBE semilocal exchange and correlation energy functional \citep{PBE}, and $ E_{\textup{x}}^{\textup{ex}} \! \left [ \left \{ \psi_{n\textbf{k}}^\sigma \right \}\right ] $ is the Hartree-Fock expression that gives the exact exchange energy of the Kohn-Sham orbitals
\begin{align}
 E_{\textup{x}}^{\textup{ex}} \! \left [ \left \{ \psi_{n\textbf{k}}^\sigma \right \}\right ] &=-\frac{1}{2} \sum_{n\textbf{k},{n}'{\textbf{k}}', \sigma } f^{\sigma }_{n\textbf{k}} f^{\sigma }_{{n}'{\textbf{k}}'}   \nonumber \\
 \times \iint_{V} \psi_{n\textbf{k}}^{\sigma\ast } \! \left ( \textbf{r} \right )  \psi_{{n}'{\textbf{k}}'}^{\sigma\ast } \! \left ( {\textbf{r}}' \right )  &v_{ee} \! \left ( \textbf{r} - {\textbf{r}}' \right ) \psi_{n\textbf{k}}^{\sigma} \! \left ( {\textbf{r}}' \right )  \psi_{{n}' {\textbf{k}}'}^{\sigma} \! \left ( \textbf{r} \right )  ,
\end{align}
where $\psi_{n\textbf{k}}^\sigma $ are the Kohn-Sham orbitals, $f^{\sigma }_{n\textbf{k}}$ are the corresponding occupation numbers. The Coulomb electron-electron interaction potential $ v_{ee} $ is defined in Eq.~(\ref{eq:LDAU_9}).

In the case of the HSE06 functional the exchange correlation energy functional has the similar form
\begin{equation} \label{eq:hybr_3}
E^{\textup{HSE06}}_{\textup{xc}}  \! \left [ \varrho,\left \{ \psi_{n\textbf{k}}^\sigma   \right \} \right ] =  E_{\textup{xc}}^{\textup{PBE}} \!  \left [  \varrho \right ]  + \alpha  E_{\textup{x}}^{\textup{ex,sr}} \! \left [ \left \{ \psi_{n\textbf{k}}^\sigma   \right \}\right ] - \alpha E_{\textup{x}}^{\textup{PBE,sr}} \!  \left [  \varrho \right ] ,
\end{equation}
where the ''sr" superscript represents the short range part of the corresponding range separated energy functional. These range separated functionals are defined via the separation of the exchange hole in the semilocal exchange functional part and the separation of the bare Coulomb interaction kernel $v_{ee}$ in the exact exchange part with a proper function of the distance $ \left|  \mathbf{r} - \mathbf{r} '  \right|$. In the HSE06 functional the range separation uses the error-function,  
\begin{equation} \label{eq:hybr_4}
  v^{\textup{sr}}_{ee} \! \left ( \textbf{r} - {\textbf{r}}' \right ) = \frac{e^{2}}{4\pi \varepsilon_{0}} \frac{1 - \textup{erf} \! \left ( \mu \left | \textbf{r} - {\textbf{r}}'   \right | \right )}{\left | \textbf{r} - {\textbf{r}}' \right |} .
\end{equation}
The expression $   v^{\textup{lr}}_{ee}  = v_{ee} -v^{\textup{sr}}_{ee} $ defines the long range part of the kernel.

For hybrid functionals one can define $ \Delta \! E^{\textup{hybrid}}_{\textup{xc}} $ to be the additional term to the semilocal PBE functional. For instance, for the PBE0 functional
\begin{align} \label{eq:hybr_6}
\Delta \! E^{\textup{PBE0}}_{\textup{xc}}  \! \left [ \varrho,\left \{ \psi_{n\textbf{k}}^\sigma   \right \} \right ]  & = E^{\textup{PBE0}}_{\textup{xc}}  \! \left [ \varrho ,\left \{ \psi_{n\textbf{k}}^\sigma   \right \} \right ] -  E_{\textup{xc}}^{\textup{PBE}} \!  \left [  \varrho \right ]   \nonumber \\
& = \alpha \left (  E_{\textup{x}}^{\textup{ex}} \! \left [ \left \{ \psi_{n\textbf{k}}^\sigma   \right \}\right ] -  E_{\textup{x}}^{\textup{PBE}} \!  \left [  \varrho  \right ]  \right ) 
\end{align}
Since  the correlation energy functional is not affected $ \Delta \! E^{\textup{PBE0}}_{\textup{xc}} = \Delta \! E^{\textup{PBE0}}_{\textup{x}} $. The corresponding non-local and orbital dependent additional potential is
\begin{align}  \label{eq:hybr_1}
\Delta \! V_{\textup{x}}^{\textup{PBE0}} \!  & \left ( \left [ \varrho \! \left ( \mathbf{r} \right ),\left \{ \psi_{n\textbf{k}}^\sigma   \right \} \right ]; \mathbf{r},{\mathbf{r}}' \right )  \nonumber \\
=\alpha & \left ( V_{\textup{x}}^{\textup{ex}} \!  \left ( \left [ \left \{ \psi_{n\textbf{k}}^\sigma   \right \}\right ]; \mathbf{r},{\mathbf{r}}' \right ) - \delta \! \left ( \textbf{r}-{\textbf{r}}' \right ) \mu_{\textup{x}}^{\textup{PBE}} \!  \left [  \varrho \! \left ( \mathbf{r} \right ) \right ]  \right ) ,
\end{align}
where the exact exchange potential is
\begin{equation} \label{eq:hybr_2}
V_{\textup{x}}^{\textup{ex}}  = - \sum_{n\mathbf{k} \sigma } f_{n\textbf{k}}^\sigma \psi_{n\textbf{k}}^\sigma\left ( \textbf{r} \right ) \psi_{n\textbf{k}}^{\sigma\ast} \left ( {\textbf{r}}' \right ) v_{ee}\left ( \textbf{r} - {\textbf{r}}' \right ) ,
\end{equation}
and $ \mu_{\textup{x}}^{\textup{PBE}} $ is the semilocal PBE exchange potential. By using the corresponding range separated Coulomb potential term in accordance with the definition of the range separated total energy terms one can similarly form the exchange potential for the HSE06 functional as well.

The above introduced total energy (Eq.~(\ref{eq:hybr_6})) and potential (Eq.~(\ref{eq:hybr_1})) can be considered as the total energy and potential correction of hybrid functionals to the semilocal potential, respectively.

\section{Connection between hybrid-DFT and DFT+$U$}

In this section we derive a connection between the description of localized states in hybrid-DFT and DFT+$U$. This connection is used to provide a theoretical foundation for the recently proposed hybrid-DFT+V$_{\textup{w}}$ method of Ref.~\onlinecite{Ivady13}.

\subsection{Effect of hybrid functionals on localized orbitals}

We begin by reformulating the additional exchange energy functional of hybrid functionals into an approximate form in order to reveal the effects of the additional term on correlated atomic-like orbitals. First we consider PBE0 in Eq.~(\ref{eq:hybr_1}), and then we discuss the case of other hybrids. The exact exchange energy, the first term on the right hand side of Eq.~(\ref{eq:hybr_6}), of a subsystem of atomic \emph{d}- or \emph{f}-like orbitals $ \phi_{m}^{\sigma} $ is defined in the last term of Eq.~(\ref{eq:LDAU_2}) as
\begin{equation} \label{eq:con_1}
E^{\textup{ex}}_{x} \! \left [ \left \{ n_{mm'}^{\sigma }  \right \} \right ] = - \frac{1}{2} \sum_{ \left \{ m \right \}, \sigma }  \left \langle m m_{1} \left | v_{ee} \right | m_{2} {m}' \right \rangle   n_{m{m}'}^{\sigma }  n_{m_{1} m_{2}}^{\sigma } .
\end{equation}
In order to determine the (semi)local PBE exchange energy of the correlated orbitals, i.e., the second term on the right hand side of Eq.~(\ref{eq:hybr_6}), we use the FLL approximation in a similar fashion as in the derivation of the DFT+$U$ method in Eq.~(\ref{eq:LDAU_3}). However, here we do not take into account the screening effect of the itinerant electrons. In this approximation the interaction energy of the localized $ \phi^{\sigma}_{m}$ orbitals can be written as \citep{LDAUrev,RinkeLDAU10}
\begin{equation} \label{eq:con_2}
E_{ee}^{\textup{DFT}} \!\! \left [ \varrho_{\textup{loc}}  \right ] \approx E_{ee}^{\textup{DFT}} \!\! \left [ n^{\sigma } \right ] = \frac{F^{0}}{2} n\left ( n-1 \right ) - \frac{J^{0}}{2} \sum_{\sigma } n^{\sigma } \left ( n^{\sigma } -1 \right ) ,
\end{equation}
where the localized density can be written as $\varrho_{\textup{loc}} = \sum_{m_{1} ,m_{2},\sigma}  \left \langle \phi^{\sigma }_{m_{1}} | \phi^{\sigma}_{m_{2}}    \right \rangle n^{\sigma }_{m_{1}m_{2}} \approx n \sum_{m, \sigma  }^{\textup{occ.}}  \left \langle \phi^{\sigma }_{m} | \phi^{\sigma}_{m}    \right \rangle  $  and $ F^{0} $ and $ J^{0} $ are the spherically averaged unscreened direct and exchange parameters of the Coulomb interaction among the localized orbitals, as defined in Eq.~(\ref{eq:LDAU_6}) and Eq.~(\ref{eq:LDAU_7}). By reformulating Eq.~(\ref{eq:con_2}) one obtains the following equation
\begin{equation} \label{eq:con_3}
E_{ee}^{\textup{DFT}} \!\! \left [ n^{\sigma } \right ] = \frac{F^{0}}{2} n^2 - \frac{F^{0} - J^{0}}{2} n - \frac{J^{0}}{2} \sum_{\sigma } \left ( n^{\sigma }  \right )^2.
\end{equation}
The first term on the right hand side of Eq.~(\ref{eq:con_3}) is the Hartree energy in the FLL approximation. This term includes the self-interaction in accordance with its definition. The rest is the (semi)local exchange energy in the FLL approximation.
By inserting Eq.~(\ref{eq:con_1}) and the appropriate part of Eq.~(\ref{eq:con_3}) into Eq.~(\ref{eq:hybr_6}) we arrive at the following form for the exchange energy correction of the PBE0 hybrid functional on the subsystem of localized atomic like orbitals:
\begin{align} \label{eq:con_4}
\Delta \! E_{x}^{\textup{PBE0}} \! \left [ \left \{ n_{mm'}^{\sigma }  \right \} \right ]   =  &- \frac{\alpha}{2}  \left ( \sum_{ \left \{ m \right \}, \sigma } \left \langle m m_{1} \left | v_{ee} \right | m_{2} {m}' \right \rangle   n_{m{m}'}^{\sigma }  n_{m_{1} m_{2}}^{\sigma }   \nonumber \right. \\ 
&- \left. \left( F^{0} - J^{0} \right) n - J^{0} \sum_{\sigma } \left ( n^{\sigma }  \right )^2 \right )
\end{align}
The corresponding additional occupation dependent potential can be obtained from the derivative of the energy functional $ \Delta \! E_{x}^{\textup{PBE0}} \! \left [ \left \{ n_{mm'}^{\sigma }  \right \} \right ]$ with respect to an element of the occupation matrix ${ n_{mm'}^{\sigma}} $, as 
\begin{align} 
\Delta \! V_{m{m}'}^{\textup{PBE0x}, \sigma}  =  &- \alpha  \left (   \sum_{m_{1}m_{2}}   \left \langle m m_{1} \left | v_{ee}  \right | m_{2} {m}' \right \rangle n_{m_{2}m_{1}}^{\sigma } \nonumber \right. \\ 
&- \left. \delta_{m{m}'}  \left (  \frac{1}{2} \left (  F^0 - J^0\right ) + J^0  n^{\sigma } \right ) \right )
\end{align}
where we have assumed that $\delta \! \varrho_{\textup{corr}} \approx \delta \! n \sum_{m, \sigma  }  \left \langle \phi^{\sigma }_{m} | \phi^{\sigma}_{m}    \right \rangle $, i.e., the infinitesimal change of the correlated charge density comes only from the variation of the on-site occupation number $ n $, so that atomic orbitals are unchanged.

In order to arrive at a more expressive form that illustrates the physical effects of the additional on-site functional term, we apply further approximations to Eq.~(\ref{eq:con_4}) and define the occupation dependent potential.  First, we just keep the matrix elements of the Coulomb matrix $  \left \langle m m_{1} \left | v_{ee}  \right | m_{2} {m}' \right \rangle $ that are only one or two center integrals 
\begin{align} 
\left \langle m m_{1} \left | v_{ee}  \right | m_{2} {m}' \right \rangle \approx  \left \langle m {m}' \left | v_{ee}  \right | m {m}' \right \rangle \delta_{mm_{2}}  \delta_{m_{1}{m}'}  \nonumber \\ + \left \langle m m_{1} \left | v_{ee}  \right | m_{1} m \right \rangle \delta_{m{m}'} \delta_{m_{1}m_{2}}  \nonumber \\
= F^{0}_{m{m}'} \delta_{mm_{2}}  \delta_{m_{1}{m}'} + J^{0}_{mm_{1}}  \delta_{m{m}'}  \delta_{m_{1}m_{2}} .
\end{align}
Using this approximation in Eq.~(\ref{eq:con_4}) gives
\begin{align} 
\Delta \! E_{x}^{\textup{PBE0}} \! \left [ \left \{ n_{mm'}^{\sigma }  \right \} \right ]   =  &- \frac{\alpha}{2}  \left ( \sum_{ m, m', \sigma }  F^{0}_{mm'}   n_{m{m}'}^{\sigma }  n_{{m}'m}^{\sigma }   \nonumber \right. \\
&+ \left. \sum_{ m \neq m_{1}, \sigma }  J^{0}_{mm_{1}}   n_{mm}^{\sigma }  n_{m_{1}m_{1}}^{\sigma }   \nonumber \right. \\
&- \left. \left( F^{0} - J^{0} \right) n - J^{0} \sum_{\sigma } \left ( n^{\sigma }  \right )^2 \right ) .
\end{align}
Similar to the approach Dudarev \emph{et al.}, we assume $ F^{0}_{m{m}'} \approx F^{0}$ and $ J^{0}_{m{m}'} \approx J^{0}$, i.e., the matrix elements are approximately equal to their mean value. Furthermore, we now choose the localized bases set $ \left \{ \phi_{m} \right \} $ in such a way that the on-site occupation matrix $ n^{\sigma}_{m{m}'}$ becomes diagonal. The result is
\begin{align} \label{eq:con_5}
\Delta \! E_{x}^{\textup{PBE0}} \! \left [ \left \{ n_{m}^{\sigma }  \right \} \right ]   =  &- \frac{\alpha}{2}  \left ( F^{0} \sum_{ m, \sigma }   \left( n_{m}^{\sigma }  \right)^{2}   
+ J^{0} \sum_{ m \neq m_{1}, \sigma }   n_{m}^{\sigma }  n_{m_{1}}^{\sigma } 
- \left( F^{0} - J^{0} \right) n - J^{0} \sum_{\sigma } \left ( n^{\sigma }  \right )^2 \right ).
\end{align}
With some additional manipulation of this expression we arrive to our main result
\begin{equation} \label{eq:con_6}
\Delta \! E_{x}^{\textup{PBE0}} \! \left [ \left \{ n_{m}^{\sigma }  \right \} \right ]   =  \frac{\alpha \left( F^{0} - J^{0} \right)}{2}  \sum_{ m, \sigma } \left (   n_{m}^{\sigma }  - \left( n_{m}^{\sigma }  \right)^{2}  \right),
\end{equation}
which describes the exchange energy correction of the subsystem of correlated orbitals for the case of the PBE0 hybrid functional. The correction potential acting on the localized atomic-like orbital $\phi_{m}^{\sigma}$ can be written as
\begin{equation} \label{eq:con_7}
\Delta \! V_{m}^{\textup{PBE0x}, \sigma}  =  \alpha  \left (  F^{0} -  J^{0} \right ) \left ( \frac{1}{2} - n_{m}^{\sigma } \right ) .
\end{equation}
We emphasize that Eqs.~(\ref{eq:con_6}) and (\ref{eq:con_7}) for hybrid functionals are the main results of this work and show a direct similarity with Eqs.~(\ref{eq:LDAU_10}) and (\ref{eq:LDAU_8}) for DFT+$U$. This similarity will be further discussed in the next subsection.

The derived formulas are strictly valid for the PBE0 hybrid functional, however, with some additional considerations we can motivate the use of the same formulas in a more general context. In the derivation, the introduction of non-local exact exchange energy functional plays the most important role and the (semi)local part has just a minor influence. The functional form of the semi-empirical B3PW91\citep{Becke93b} and B3LYP \citep{B3LYP} hybrid functionals differ from the PBE0 functional in the semilocal DFT part only. Therefore, if we simply assume that the FLL approximation in Eq.~(\ref{eq:con_2}) is roughly valid for the more complex expression of the semilocal part of these two functionals, then the final result apply to them as well.

In the case of the range separated HSE06 functional, defined in Eq.~(\ref{eq:hybr_3}), the electron-electron interaction potential $v_{ee}$ is separated in space in accordance with Eq.~(\ref{eq:hybr_4}). In our derivation, this new potential enters into the formulas of the definition of the unscreened parameters of the Coulomb interaction (Eq.~(\ref{eq:LDAU_6})). Without the calculation of these integrals one can immediately see that $ \tilde{F}^{0}_{m{m}'} \! \left  ( \mu  \right ) < F^{0}_{m{m}'} $ if $ 1/\mu \neq \infty $. The considered states $ \left \{ \phi_{m} \right \} $ are well localized, for $3d$-orbitals the maximal distance of the electron density maxima is 1--2~\AA, while the cut-off radius is typically $\mu \approx 5$~\AA. Therefore, the assumption $  \tilde{F^{0}}_{m{m}'}  \approx F^{0}_{m{m}'} $ is reasonable.

\subsection{The hybrid-DFT+V$_{\textup{w}}$ method}

As was concluded in the derivation of Eqs.~(\ref{eq:con_6}) and (\ref{eq:con_7}) there is a direct correspondence between the energy and potential in the hybrid scheme and in the formulation of DFT+$U$ by Dudarev \emph{et al.}\ in Eqs.~(\ref{eq:LDAU_10}) and (\ref{eq:LDAU_8}).  However, the strength of the on-site interaction potential is defined in different ways. In the optimal case, the potential strengths would be equal to the strength of the on-site potential in the real system. In the DFT+$U$ method this is formally represented by the definition
\begin{equation} 
 U^{\textup{DFT+U}}_{\textup{eff}} = U - J = U^{\textup{real}}_{\textup{eff}},
\end{equation}
On the other hand, in hybrid-DFT the following equation needs to be satisfied
\begin{equation} \label{eq:par_1}
 U^{\textup{hybrid}}_{\textup{eff}} = \alpha  \left (  F^{0} -  J^{0} \right )  = U^{\textup{real}}_{\textup{eff}}.
\end{equation}
This expression shows that the mixing parameter $\alpha$ in hybrid functionals determines the strength of the screening of the bare on-site Coulomb interaction. This mixing parameter thus needs to be chosen properly to reproduce the desired potential strength. 

Despite the equivalent effect of the two methods on localized orbitals, still there are significant differences. In DFT+$U$ method a selected subset of correlated states are affected, the interaction among the delocalized states and delocalized and correlated states are described on the basis of (semi)local DFT . Nevertheless, this method allows the use of different $U_{\textup{eff}} $ for different orbitals or atoms. In contrast, in hybrid functionals all the electron-electron exchange and correlation effects are subject to an equivalent treatment governed by the mixing parameter. In other words, the use of $\alpha < 1$ gives a homogeneous and global screening of the electron-electron interaction in the system. 

In transition metal (TM) oxides (TMOs) or in other TM compounds states related both to $sp^{3}$ hybridization and to $d$-orbitals are present simultaneously. It cannot be generally expected that the same screening is suitable for these different states. Therefore, within the usual hybrid-DFT scheme, such correlated systems can not be faithfully described. However, this description can still be better than in DFT+$U$, since the $sp^{3}$ states may be treated better. In the case of localized states the bare on-site parameters $F^{0} $ and $ J^{0} $ are large, i.e., a few tens of eV. A small deviation in $\alpha$ can therefore result in a large increase or decrease of the on-site interaction strength. The fact that the effect of the deviation in $\alpha$ on $sp^{3}$ states is smaller due to the weaker bare interaction between the less localized orbitals, suggests that an $\alpha$ that fulfils Eq.~(\ref{eq:par_1}) can be a good choice for correlated semiconducting TM compounds. Hence, as pointed out by Perdew \emph{et al.}\citep{Perdew07}\ the global and homogeneous screening approximation in the hybrid-DFT scheme is rather limiting. To overcome this issue, space, orbital or energy dependent mixing parameter have been proposed. On the other hand, resting on the fact that hybrid-DFT and DFT+$U$ methods introduce the same correction on the subsystem of localized orbitals, a combination of these two methods can bring advantages over using them separately. 

On this basis, we suggested, in a previous work \citep{Ivady13}, the hybrid-DFT+V$_{\textup{w}}$ scheme. It introduces an additional on-site screening potential
\begin{equation} \label{eq:par_5}
 V_{m}^{ \sigma} \! \left( w \right)  =  w \left ( \frac{1}{2}  - n_{m}^{\sigma }  \right ) 
\end{equation}
to a subset of localized orbitals in a \emph{hybrid functional}. This potential can be obtained from the derivative of the total energy expression
\begin{equation} \label{eq:par_3}
 \Delta E \! \left( w \right) =  \frac{  w  } {2} \sum_{m, \sigma} \left(  n^{\sigma}_{m} - {\left( n^{\sigma}_{m} \right)}^{2} \right) .
\end{equation}
The strength of the additional correction and potential is defined as
\begin{equation} \label{eq:par_4}
w = - \left( U^{\textup{hybrid}}_{\textup{eff}} - U^{\textup{real}}_{\textup{eff}} \right).
\end{equation}
In contrast with DFT+$U$ and hybrid-DFT methods this scheme allows for the additional degrees of freedom to describe both $sp^{3}$ hybridized and $d$-orbital related states. A further practical advantage is that the aforementioned two methods are quite popular and often implemented in first principles codes in such a way that they can be used simultaneously, which allows the use of the hybrid-DFT+V$_{\textup{w}}$ scheme with no need for further implementation. 

\subsection{Self-consistent determination of parameter $w$}

A practical scheme to satisfy Eq.~(\ref{eq:par_4}) was demonstrated in Ref.~\onlinecite{Ivady13} where we determined the strength of the on-site correction potential $w$  by the fulfillment of the ionization potential (IP) theorem \citep{Perdew_ex, Perdew97,Almbladh85} or, in other context, the generalized Koopmans' theorem\citep{LanyZunger09,LanyZunger10,Dabo} (gKT). These theorems state that the KS eigenvalue of the highest occupied KS orbital is equal to the negative ionization energy and remains constant under the variation of its occupation number in the case of the exact exchange correlation functional. For approximate density functionals the IP theorem is usually not upheld with satisfactory accuracy. On the other hand, construction of exchange correlation functionals that possess the above mentioned criteria have been generally successful\citep{Cococcioni05,Cococcioni06,LanyZunger09,LanyZunger10,Dabo,Baer10,Xiao11,Gaiduk12,Baer10,Srebro12,KronikJCTC12, KronikPRL12,KronikPRB13,KummelJCP13,Atalla13,Richter13, Kronik13}. The degree to which a functional upholds the IP theorem or the gKT can be checked via the non-Koopmans' energy \citep{Dabo}, which is the difference of the KS eigenvalue of the highest occupied orbital and the negative ionization energy in the external potential $v_\textup{ext} \! \left( \mathbf{r} \right)$. Despite the arbitrary constant shift of the KS potential in periodic systems, which makes the single particle energies physically meaningless, the non-Koopmans' energy can still be well defined. However, in charged periodic systems the KS eigenvalues and total energies are additionally shifted due to the spurious electrostatic interaction of the localized charge density with its periodically repeated images and with the neutralizing jellium background. These effects are due to the periodic supercell approximation and should be eliminated from the non-Koopmans' energy using
\begin{equation} \label{eq:NK}
  E^{\textup{NK}}_{i} = \left( \varepsilon_{i}  + \delta \varepsilon^{\textup{cc}}_{i,q} \right)  - \Delta E_{N},
\end{equation}  
where $\varepsilon_{i}$ can be either the highest occupied or the lowest unoccupied KS eigenvalue in the system of either $N$ or $N-1$ electrons, respectively,  $\delta \varepsilon^{\textup{cc}}_{i,q} $ is the charge correction of the KS orbital in the corresponding charged state $q$ and
\begin{equation} \label{eq:me_2}
\Delta E_{N} = \left( E_{N}  +  \delta E^{cc}_{q} \right) - \left( E_{N-1} +  \delta E^{cc}_{q-1} \right),
\end{equation}
 where $ E_{N}$ is the total energy of the system of $N$ electrons and  $\delta E^{cc}_{q}$ is the charge correction of the total energy, where the charge state $q = N - N_{0}$ and the $N_{0}$ is the number of electrons in the neutral system. In accordance with the IP theorem the KS eigenvalue of the highest occupied orbital is constant during the occupation, therefore
\begin{equation} \label{eq:me_3}
 \Delta \varepsilon_{i} = \left( \varepsilon^{\textup{occ}}_{i}  + \delta \varepsilon^{\textup{cc}}_{i,q} \right) - \left( \varepsilon^{\textup{unocc}}_{i}  + \delta \varepsilon^{\textup{cc}}_{i,q-1} \right).
\end{equation} 
This quantity may indicate the same error as the Non-Koopmans' energy. The condition of  $ E^{\textup{NK}}_{i} = 0 $ and  $\Delta \varepsilon_{i} = 0$, i.e., the fulfillment of the IP theorem or gKT, may present a more precise self-interaction free description of the orbitals \citep{Dabo}. 

\subsection{The band gap in DFT+$U$ and hybrid-DFT}

Our connection between DFT+$U$ and hybrid functionals can also be used to better understand the effect these two theories have on the band gap. Following the work of Gr\"uning \emph{et.~al.}\citep{Gruning06}\ the derivative discontinuity is the discrepancy between the KS gap and the real or quasi-particle (QP) gap, 
\begin{equation}
\Delta_{\textup{dd}} = \varepsilon_{\textup{gap}}^{\textup{QP}} - \varepsilon_{\textup{gap}}^{\textup{KS}}.
\end{equation}
In accordance with many-body perturbation theory, we define the quasi-particle energies from the KS eigenvalues $\varepsilon_{i}^{\textup{KS}} $ and orbitals $\psi_{i}$ as
\begin{equation}
\varepsilon_{i}^{\textup{QP}} \approx \varepsilon_{i}^{\textup{KS}} + \left \langle \psi_{i} \left |  \Sigma \! \left (  \varepsilon^{\textup{QP}}_{i} \big/ \hbar \right ) - \mu_{\textup{xc}} \right | \psi_{i} \right \rangle,
\end{equation}
where $\mu_{\textup{xc}} $ is the (semi)local exchange correlation potential. If we approximate the non-hermitian and energy dependent self-energy $\Sigma \! \left (  \varepsilon_{i} \big/ \hbar  \right )$ with the hybrid exchange correlation potential, the derivative discontinuity introduced by the hybrid functional is 
\begin{equation}
\Delta_{\textup{dd}} = \left \langle \psi_{i+1} \left |  \Delta V^{\textup{hyb}}_{\textup{xc}} \right | \psi_{i+1} \right \rangle - \left \langle \psi_{i} \left |  \Delta V^{\textup{hyb}}_{\textup{xc}} \right | \psi_{i} \right \rangle,
\end{equation}
where $\Delta V^{\textup{hyb}}_{\textup{xc}} $ can take the form of Eq.~(\ref{eq:hybr_1}), for instance. The matrix elements can be calculated using $\psi_{i+1} = \phi^{\textup{unocc}}_{{m}'}$ and  $\psi_{i} = \phi^{\textup{occ}}_{m}$, giving
\begin{equation}
\Delta_{\textup{dd}} = \alpha \left( F^{0} - J^{0} \right),
\end{equation}
which thus is equal to the strength of the potential introduced in hybrids (Eq.~(\ref{eq:con_7})). Similarly, for DFT+$U$ method one obtains $\Delta_{\textup{dd}} = U_{\textup{eff}}$\citep{LDAUrev}. Hence, for the case of localized atomic like orbitals the introduction of the approximations used in hybrid-DFT and DFT+$U$ method, with a correct $U_{\textup{eff}}$ parameter, introduces a derivative discontinuity between the occupied and unoccupied orbitals of the magnitude of the potential.


\section{Application of the HSE06+V$_\textup{w}$ scheme}

In this section we demonstrate the necessity of the application and the use of hybrid-DFT+V$_\textup{w}$ method on the system of substitutional chromium (Cr$_{\textup{Al}}$) in wurtzite AlN and substitutional vanadium (V$_{\textup{Si}}$) in 4H-SiC. The effect of the additional correction potential on the electronic structure and on the localization of KS orbitals and the spin density are thoroughly investigated.


\subsection{Methodology}

Our calculations use DFT in a plane wave basis set in the PAW \citep{PAW,Kresse99} formalism as implemented in the 5.3.3 version of  Vienna Ab Initio Simulation Package (VASP)\cite{VASP, VASP2}. To model isolated defects, large supercells of 578 and 432 atom are used for 4H-SiC and wurtzite AlN (w-AlN), where the vanadium and the chromium impurity are embedded on the silicon and aluminum site, respectively. The supercell is big enough for $\Gamma$-point sampling of the Brillouin zone to be sufficient for obtaining a convergence.

The electronic and structural parameters of the 4H-SiC are well reproduced with HSE06 functional\citep{Kresse06,Kresse08}.  However, to improve the correspondence between the KS quasi-particle gap and the experimental gap of w-AlN we slightly modify the parameter set of HSE06 functional for this system. In the case of range separated hybrids (Eqs.~(\ref{eq:hybr_3}) and (\ref{eq:hybr_4})) both the mixing parameter $\alpha$ and the range separation parameter $\mu$ are related to the predicted band gap of semiconductors\citep{Pasquarello10}. The former one is additionally connected to the description of local physics\citep{KronikPRL12}, therefore it affects the predicted lattice constants as well. With the HSE06 functional these parameters are well reproduced for w-AlN. The deviation from the experimental value\citep{book_AlN} is $0.2$\% and $0.04$\% for parameter $a$ and $c/a$, respectively. These results suggest that the original $\alpha = 0.25$ setting is suitable for this material. Additionally, the mixing parameter $\alpha$ is connected to dielectric constant $\varepsilon$ of semiconductors\citep{Iori12,Marques11}. The fact that 4H-SiC and w-AlN have similar dielectric constant also supports the use of the original mixing parameter.
On the other hand, the band gap of w-AlN is underestimated in HSE06 calculation, $E^{\textup{HSE06}}_{\textup{gap}} =5.65$~eV while the experimental value \citep{ALN_gap} at room and zero temperature is $E^{\textup{exp}}_{g} = 6.03$~eV and $6.12$~eV, respectively. With the choice of $0.1$~\AA$^{-1}$ for the new value of the range separation parameter $\mu$, from now we refer to this functional as mHSE, we can preserve the accuracy of the predicted lattice parameters, $a = 3.1030$~\AA\ and $c/a =1.6018$ with $0.3$\% and $0.06$\% deviation from experimental values, respectively, and improve the KS quasi-particle gap, $E^{\textup{mHSE}}_{\textup{gap}} = 5.96$, on the cost of reasonable increment of computational time. 

In order to evaluate the non-Koopmans' energy (Eq.~(\ref{eq:NK})) in our periodic supercells properly we fix the geometry during the examination of the exactness of the functional and use charge correction to eliminate the spurious electrostatic interaction of charged point defects with their periodically repeated images and with the compensating homogeneous charge distribution. Here we used the HSE06 or mHSE relaxed geometry of the system under consideration where the examined KS orbital $i$ is the highest occupied one. On the other hand, the issue of charge correction is a long standing problem of point defect calculations in periodic codes \citep{LanyZunger08,Dabo08,Pasquarello12}. Even though today there are relatively reliable correction schemes for the total energy correction \citep{Pasquarello12}, the correction of KS orbitals is still not generally well defined. In our examples we applied the following strategy: For the charge correction of the total energy we used the correction scheme introduced by Freysoldt \emph{et.~al.} \citep{Freysoldt}, which works well for a localized charge distribution \citep{Freysoldt, Pasquarello12}. To avoid the correction of the KS orbitals, we considered only the highest occupied ones in the neutral charge state of the defects to evaluate the quality of the functional form. 

As we demonstrated earlier \citep{Ivady13}, the failure of the hybrid functional can be remedied within the hybrid-DFT+V$_{\textup{w}}$ scheme by the correction of Eq.~(\ref{eq:par_5}). In practice the VASP code uses the approach of Dudarev \emph{et al.}\ of the DFT+$U$ method, which provides us the desired potential correction form. In this method the parameter of the potential, $U_{\textup{eff}} = U - J$, represents the strength of the on-site screened effective interaction potential. In our formalism, however, the parameter of the potential is $w$, which has a different meaning in accordance to Eq.~(\ref{eq:par_4}). This parameter may take both positive and negative values and can be determined self-consistently by enforcing the fulfillment of the generalized Koopmans' theorem, $E^{\textup{NK}}_{\textup{ho}} = 0$, via the non-Koopmans' energy as outlined above.

For calculating the hyperfine (hf) constants\citep{Blochl00,Szasz13} we used a plane wave cut-off of 420 eV, which was sufficient  to obtain convergent spin density and hf constants. Recently it was shown \citep{Szasz13} that for the calculation of the hf constants related to point defects in semiconductors, e.g. in SiC, the HSE06 functional provides the accurate results with taking the contribution of the spin polarization of the core electrons to the Fermi-contact term into account.


\subsection{Cr$_{\textup{Al}}$ in w-AlN}

\begin{figure}
\includegraphics[width=0.7\columnwidth]{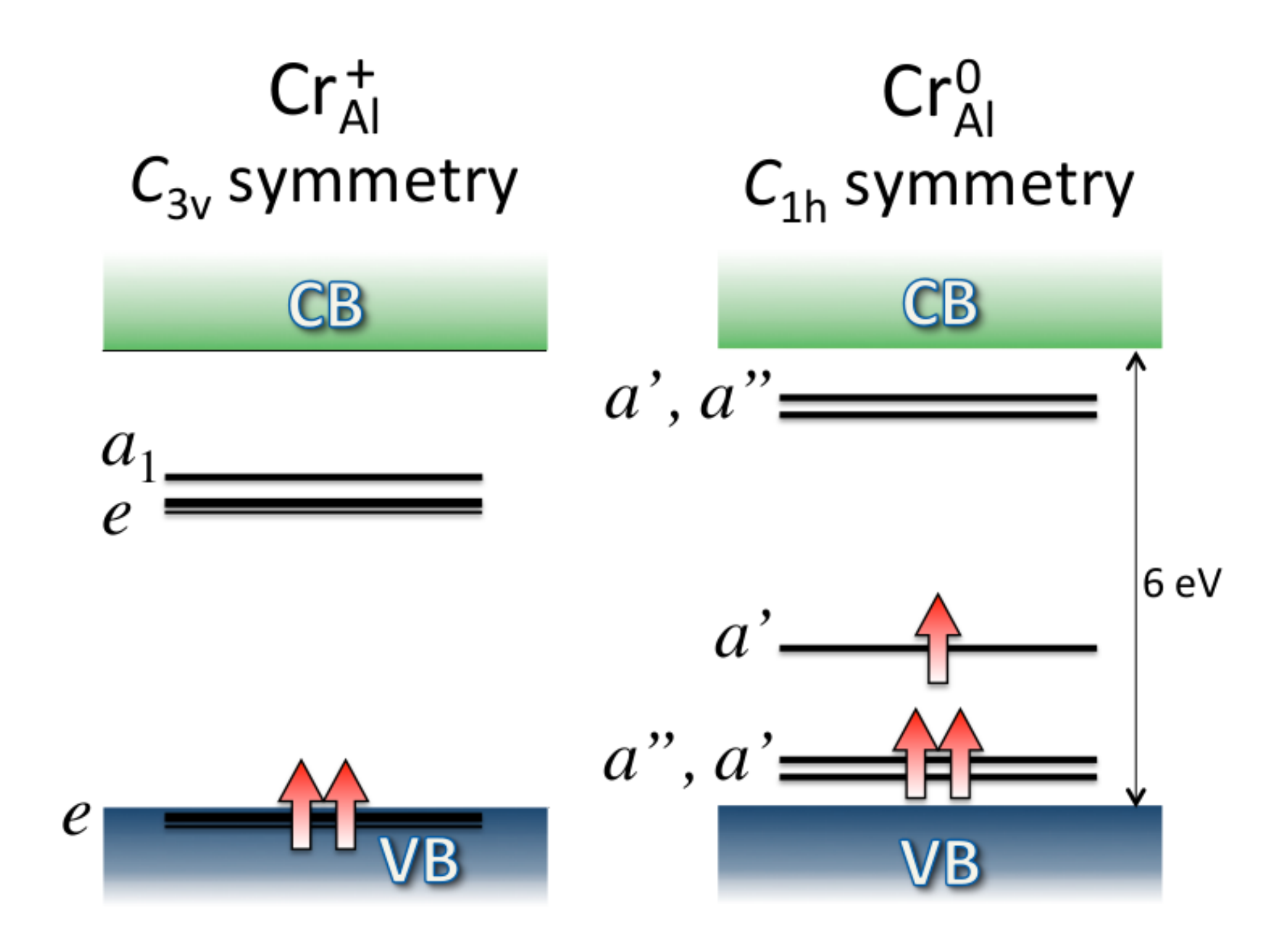}
\caption{\label{fig:Cr_lev}(Color online)  Schematic diagram of the defect orbitals of the positively charged and the neutral Cr$_{\textup{Al}}$ point defect in w-AlN. }
\end{figure}

\begin{figure}
\includegraphics[width=1.0\columnwidth]{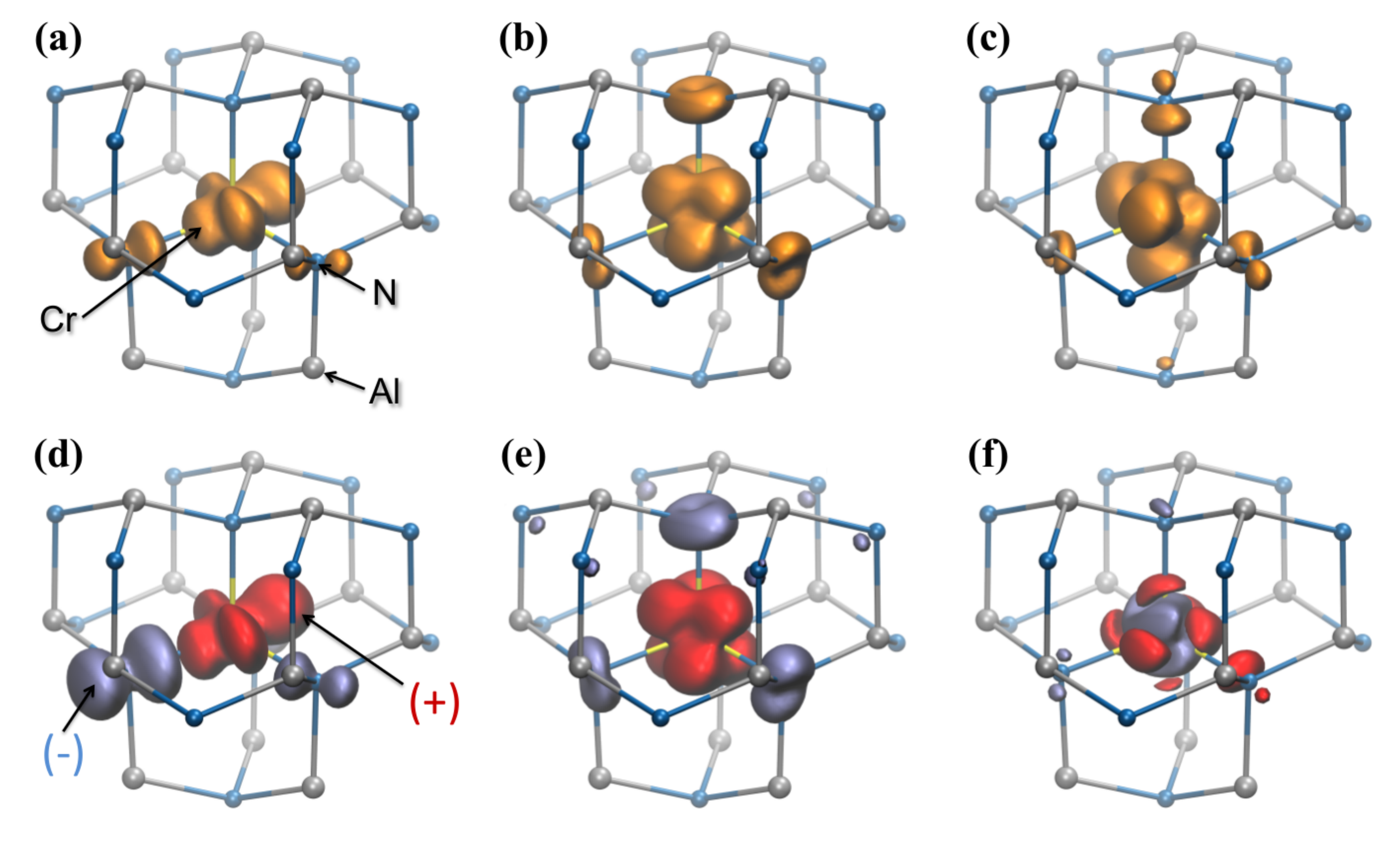}
\caption{\label{fig:Cr_chg}(Color online) Single particle charge densities and their change due to the correction of the functional. The charge density of the gap states of the neutral Cr$_{\textup{Al}}$ point defect in w-AlN (see Fig.~\ref{fig:Cr_lev}) are shown in the upper part of the figure, i.e., (a) the highest occupied defect orbital $a'$,  (b) the occupied lower lying split $e$ defect orbital and (c) the lowest unoccupied split  $e$ defect orbital as calculated with mHSE method and plotted with the isosurface value is 0.05. The following  (d), (e) and (f) figures show the change of the electron density of the corresponding defect orbitals due to the correction V$_{\textup{w}}$ of the mHSE hybrid functional with $w=-1.6$ (see text for more explanation). The isosurface values are 0.005, 0.01 and 0.0005, respectively. The red (dark grey) and blue (light grey) lobes indicate increased and decreased localization, respectively.}
\end{figure}

First, we present and discuss the electronic structure of the substitutional Cr$_{\textup{Al}}$ point defect based on the tight binding picture of the orbitals, group theory considerations and the results of mHSE calculations. The schematic diagram of the impurity related KS orbitals is shown in Fig.~\ref{fig:Cr_lev} for two different charge states. In the highest C$_{\textup{3v}}$ symmetry of the hexagonal supercell with a defect the five times degenerate atomic $d$-orbital of the Cr splits into two $e$ and one $a_{1}$ states, as can be seen in the case of positively charged Cr$_{\textup{Al}}$. The higher lying $e$ state and the $a_{1}$ states above originate from the three times degenerate $t_{\textup{2}}$ orbital splits due to the hexagonal crystal field of about 0.25 eV. In the lower symmetry of C$_{\textup{1h}}$ the highest dimension of the irreducible representations is one, therefore the double degenerate states split into $a'$ and $a''$ states and the $a_{1}$ state transforms into $a'$ state.  On the other hand, the four dangling bonds of the neighbor nitrogen atoms form one $e$ and two $a_{1}$ states in C$_{\textup{3v}}$ symmetry and an $a''$ and three $a'$ states in C$_{\textup{1h}}$ symmetry. These vacancy states are originally occupied with five electrons, however, driven by the large difference of the electron negativity of N and Cr they capture three further electrons from the Cr atom to get fully occupied. In the neutral charge state of the point defect the Cr impurity can be considered as Cr$^{\textup{3+}}$. Nevertheless, the atomic like states and the vacancy states belong to the same irreducible representation and can mix with each other. As a result, the realized impurity states are never pure $d$-like states, and there is a finite localization on the neighbor N atoms as can be seen on the top part of Fig.~\ref{fig:Cr_chg}. Additionally, neither of the vacancy related states are a pure mixture of the dangling bonds (not shown). These later orbitals are found deeply into the valence band, while most of the impurity states appear in the large band gap of w-AlN (Fig.~\ref{fig:Cr_lev}), and are occupied by three and two electrons with parallel spins in the neutral and the positively charged states, respectively. 

\begin{figure}
\includegraphics[width=0.8\columnwidth]{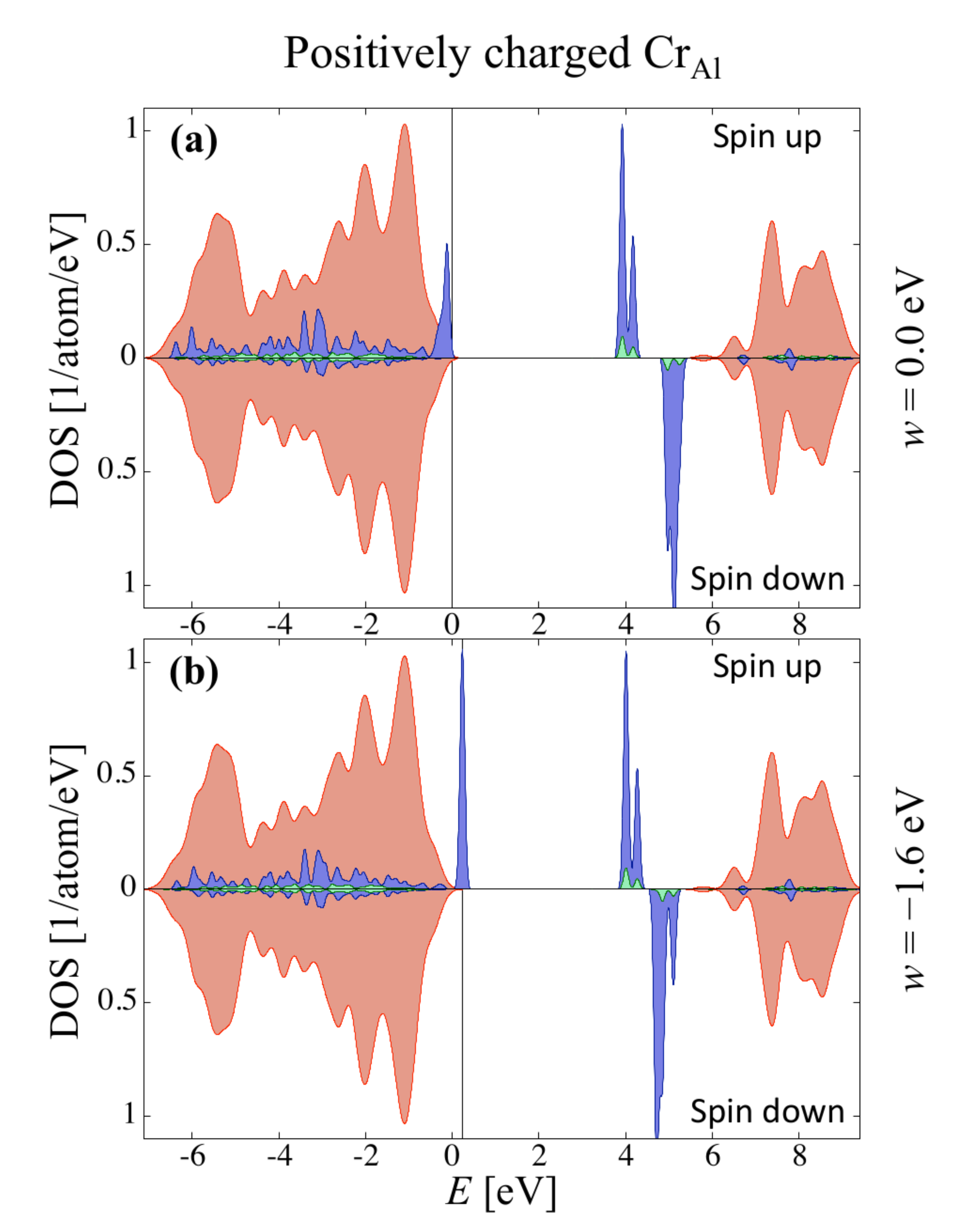}
\caption{\label{fig:pdos_Cr_p}  (Color online) Total and partial density of the states of the host w-AlN and the Cr impurity in the positively charged Cr$_{\textup{Al}}$ point defect, respectively. The red filled curves show the total DOS of the host, while blue and green filled curves show the $d$ and $sp$ partial DOS of the Cr.  These later curves were scaled up to be visible. Figure (a) and (b) show the results of the calculations with mHSE and mHSE+V$_{\textup{w}}$ exchange correlation functional (see text for more explanation).}
\end{figure} 

\begin{figure}
\includegraphics[width=0.8\columnwidth]{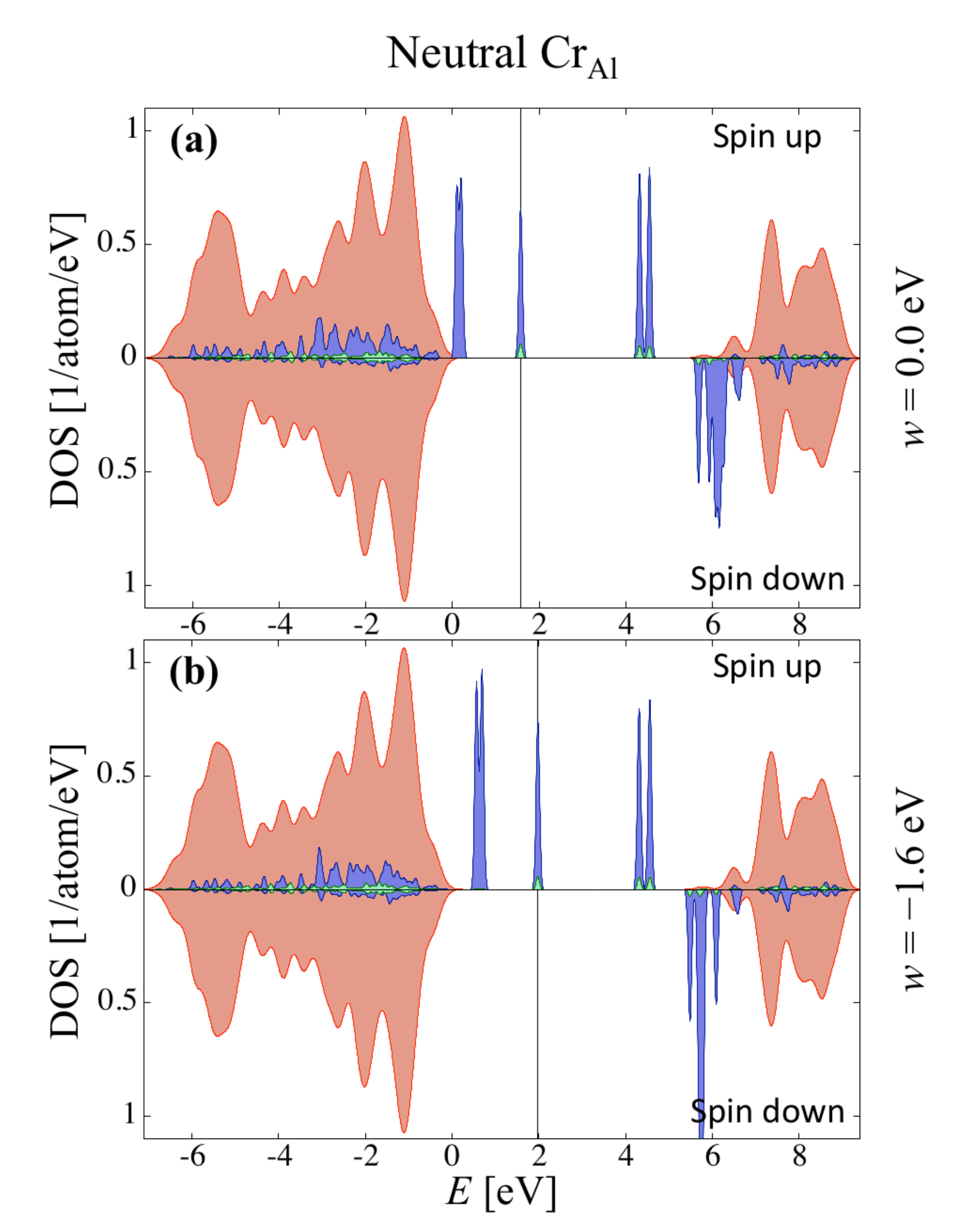}
\caption{\label{fig:pdos_Cr_0}  (Color online) Total and partial density of the states of the host w-AlN and the Cr impurity in the neutral Cr$_{\textup{Al}}$ point defect, respectively. The red filled curves show the total DOS of the host, while blue and green filled curves show the $d$ and $sp$ partial DOS of the Cr.  These later curves were scaled up to be visible. Figure (a) and (b) show the results of the calculations with mHSE and mHSE+V$_{\textup{w}}$ exchange correlation functional (see text for more explanation).}
\end{figure} 

The partial density of the state (pDOS) plot of the impurity states can be seen in Figs.~\ref{fig:pdos_Cr_p}-\ref{fig:pdos_Cr_0}. One might notice that according to the result of the mHSE calculation there is no double positive charge state, because in the positive charge state of Cr$_{\textup{Al}}$ all of the occupied defect states fall into the valence band. On the other hand there are experimental indications\citep{Gerstmann00,Gerstmann01} and theoretical predictions \citep{Baur95} of the existence of Cr$^{\textup{5+}}$ in w-AlN. This contradiction may indicate the inaccuracy of the mHSE functional and the necessity of the correction. Here, we would like to mention that the applied modification in the parameter set ofthe  HSE06 functional lowers the valence band edge\citep{Pasquarello10} with approximately $0.2$~eV in the case of mHSE. Without this modification the $e$ state falls deeper into the valence band and as a consequence enhanced error is expected in the case of HSE06 functional. 

In order to examine the accuracy of the description of the highest occupied localized orbital we have calculated the non-Koopmans' energy in accordance with Eq.~(\ref{eq:NK}). In the evaluation of this quantity we have to restrict the calculations to the fix C$_{\textup{1h}}$ geometry of the neutral charge state. The charge correction of the total energy in the positively charged state of Cr$_{\textup{Al}}$ was $\delta E^{\textup{cc}}_{+} = 0.18$~eV. The determined non-zero value of the $E_{\textup{NK}} = -0.14$~eV which indicates that the treatment of this orbital is not faithful in the mHSE method. 

\begin{figure}
\includegraphics[width=0.85\columnwidth]{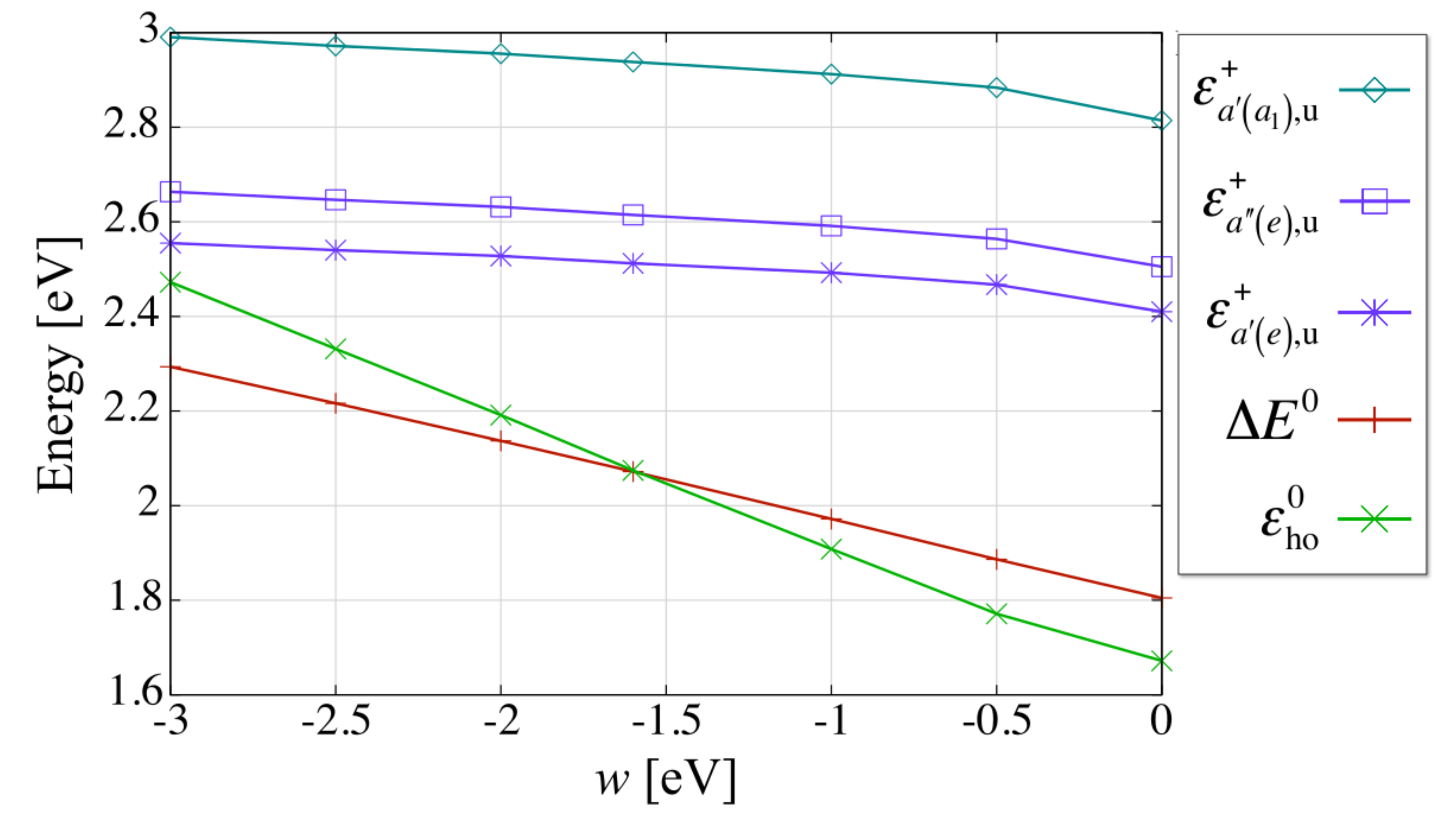}
\caption{\label{fig:Cr_NK} Variation of Kohn-Sham (KS) eigenvalues and the charge corrected total energy difference (see text for more explanation) with respect to the strength of the correction parameter $w$ of mHSE+V$_{\textup{w}}$ method in the case Cr$_{\textup{Al}}$ in w-AlN. The variance of the highest occupied KS orbital in the neutral charge states and the unoccupied states in the positively charged state are shown as obtained on the fix geometry of the neutral state. The total energy difference is calculated from the total energies of the two charge states  with applied charge correction. The valence band edge is chosen to possess the zero value on the energy scale. }
\end{figure}

In order to determine the parameter of the correction potential V$_{\textup{w}}$ we consider the variation of the important KS orbitals as well as the total energy difference, Eq.~(\ref{eq:me_2}), with respect to the strength of the potential $w$ as shown in Fig.~\ref{fig:Cr_NK}. Interestingly, not just the KS energy of the highest occupied states $\varepsilon^{0}_{\textup{ho}}$ , but the total energy difference $\Delta E^{0}$ decreases rapidly with the variation of parameter $w$, which may indicate qualitative changes in the description of this orbital. As a consequence of the similar slope of these linear curves, the relatively small $E_{\textup{NK}}$ can be eliminated only with a relatively large correction potential.  It can be seen in Fig.~\ref{fig:Cr_NK} that the two curves of $\varepsilon^{0}_{\textup{ho}}$ and $\Delta E^{0}$ cross each other at $w=-1.6$~eV, which is the strength of the needed correction potential to fulfill the generalized Koopmans' condition. To fulfill Eq.~(\ref{eq:me_3}) one needs $\delta \varepsilon^{\textup{cc}}_{+} =-0.44$~eV charge correction for the lowest unoccupied KS orbitals in the positive charge state.  

The application of the correction V$_{\textup{w}}$ shifts the energy upward of both the KS orbitals and the $(+|0)$ charge transition level  by $0.4$~eV and $0.27$~eV, respectively. The consequence is that a double positive charge state is predicted (see Fig.~\ref{fig:pdos_Cr_p}). 

\begin{figure}
\includegraphics[width=0.95\columnwidth]{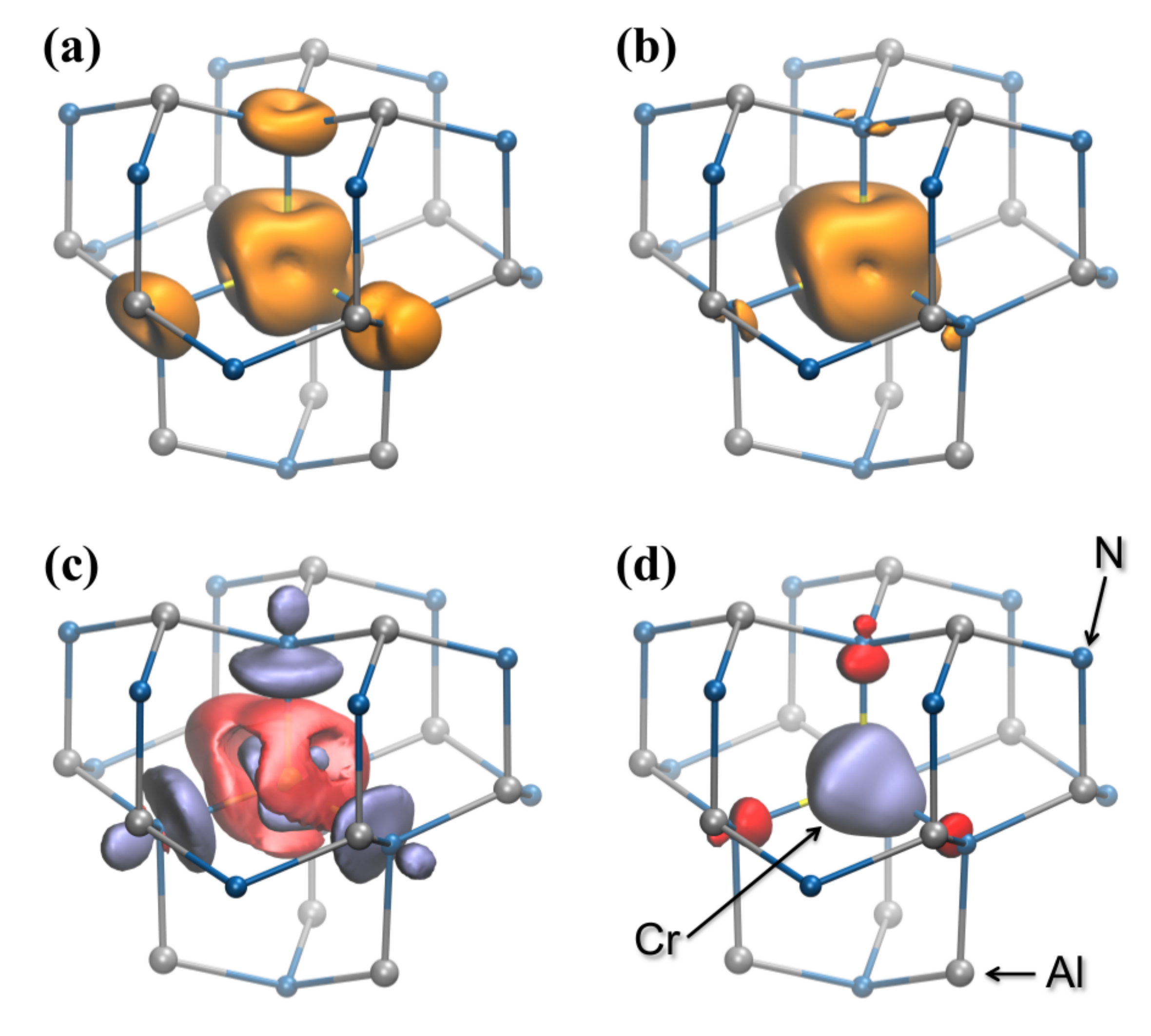}
\caption{\label{fig:Cr_chg3} (Color online) Change of the total charge density and the spin density upon the correction of the hybrid functional mHSE. (a) The summed charge density of the occupied KS orbitals in the band gap and (b) the spin density of the Cr$_{\textup{Al}}$ in w-AlN are shown with the isosurface value of 0.05.  Figure (c) presents the change of the total charge density while figure (d) the change of the spin density as a response to the additional potential V$_{\textup{w}}$ with $w=-1.6$~eV. In both cases the red (dark grey) and blue (light grey) lobes represent increase and decrease of the density, respectively. The isosurface value in (c) and (d) were chosen to 0.002 and 0.01, respectively. }
\end{figure} 

To further investigate the influence of the $w$ parameter shown in Fig.~\ref{fig:Cr_NK}, we studied the change of the physically measurable quantities such as the total charge density and the spin density (Fig.~\ref{fig:Cr_chg3}) and the charge density distribution of the localized $d$-like orbitals and their changes due to the correction V$_\textup{w}$ (Fig.~\ref{fig:Cr_chg}). The effect of the correction in the case of negative $w$ parameter is self repulsion, see Eq.~(\ref{eq:par_5}), which makes the atomic $d$-orbitals less favourable and suggests decreased localization. In the case of the total charge density the delocalization occurs only in the region of the largest value of charge densities and get localized in the neighbor shells, while larger and continuous delocalization can be observed at the Cr site for the spin density. In both cases there are contributions from the neighboring N atoms. On the other hand, interestingly, the $d$-like orbitals in the band gap get more and more localized while the localization on the dangling bonds of the neighbor N atoms decreases (see Fig.~\ref{fig:Cr_chg}) which is unexpected.

In order to quantify the effect of the correction V$_{\textup{w}}$ on the electron density we calculated the hyperfine tensor with the mHSE and mHSE+V$_{\textup{w}}$ functionals, the results are shown in  Table~\ref{tab:hyperfine}. The hyperfine tensor is related to the degree of localization of the spin density on the atoms. In the case of Cr$_{\textup{Al}}$ point defect, the hyperfine matrix elements decrease due to the applied correction, as it is expected, however the magnitude of the change is a fraction of the total splitting.  

\begin{table}
\begin{ruledtabular}
\caption{\label{tab:Cr_tot} Projected on-site charge density and spin polarization of Cr impurity at Al site of w-AlN in its neutral charge state. The total, $sp$ and $d$ projected occupations as well as the $d$ projected occupation of the gap states are presented for the cases of mHSE and mHSE+V$_{}\textup{w}$, $w = -1.6$, functionals. The applied PAW potential included the fully occupied $3p$ orbitals as well. The projection of the KS orbitals onto the atomic orbitals of the Cr was carried out inside the integration sphere of $r_{\textup{WZ}} = 1.323 $~\AA .}
 \begin{tabular}{ c|ccc|c }
  Projected on-site       & ~       & ~      & ~      & ~                                                   \\
  charge density of Cr  & Total & $sp$ & $d$  & $d^{\textup{occ.}}_{\textup{gap}}$ \\
  \hline
  mHSE	& 10.855 &  6.775 & 4.080 & 1.419\\
  mHSE+V$_{\textup{w}}$ & 10.876 &  6.764 & 4.112 &  1.752\\ \hline
  $\Delta$	& 0.021   & -0.011 & 0.032 &  0.333\\ \hline \hline
   Projected on-site      & ~       & ~      & ~      & ~                                                   \\
   magnetization of Cr  & Total & $sp$ & $d$  & $d^{\textup{occ.}}_{\textup{gap}}$ \\
  \hline
  mHSE	&  2.781 &  0.064 & 2.718  & 1.419\\
  mHSE+V$_{\textup{w}}$ &  2.684 &  0.064 & 2.621  &  1.752\\ \hline
  $\Delta$	&-0.097 &  0.000 & -0.097  & 0.333\\
 \end{tabular}
\end{ruledtabular}
\end{table}

To explain the observed opposite behavior of the total density and the density of the localized defect orbitals we have to recall that the defect states are not pure $d$-like or host related vacancy orbitals, but, as we mentioned earlier, they are linear combinations of both. This can be observed in the partial density of states (Figs.~\ref{fig:pdos_Cr_p}-\ref{fig:pdos_Cr_0}) as well as in the charge density of the localized orbitals (Fig.~\ref{fig:Cr_chg}). There is a large charge and spin density localization on the $d$-orbitals of the Cr atom coming from the states of the valence band. These are quantified in Table~\ref{tab:Cr_tot} with the integrated projections. As can be seen, the change of the localization of the gap states is approximately an order of magnitude larger than the change of the total and spin density localization on the Cr atom. It is only possible if the valence band related states undergo an opposite change, i.e., the  localization on the $d$-orbitals largely decreases, while on the vacancy related orbitals it increases. The sum of the large but opposite response of the gap states and the valence related states gives the change of the total and spin density (Fig.~\ref{fig:Cr_chg3}). It thus appears the observed behavior is due to the applied correction counteracting the formation of linear combinations of the atomic $d$-like states and the vacancy related  $sp$-orbitals, i.e., it makes the impurity states more atomic like and the host related states more host related.
It is possible that the result is an increase of the KS energies of the occupied states and the total energy, which may explain the decrease of the total energy difference and the KS energy of the unoccupied orbitals in response to increasing strength of the correction potential.

Hence, in summary increased V$_{\textup{w}}$ decreases the localization of the highly localized part of the $d$-orbitals, and rearranges the system of KS particles to form less mixed impurity and valence states.  
This means that the mHSE hybrid functional over localize the correlated states and overestimate the contribution of the orbitals for the valence band states.

\begin{table}
\begin{ruledtabular}
\caption{\label{tab:hyperfine}  Comparison of the calculated and measured hyperfine parameters of Cr$_{\textup{Al}}$ in w-AlN and V$_{\textup{Si}}$ in 4H-SiC. } 
 \begin{tabular}{ c|ccc }
   Cr$_{\textup{Al}}$ in w-AlN & $A_{\parallel }$ [MHz] & $A_{\perp}$ [MHz]  \\
  \hline
  mHSE	& 13.0 & 26.9 \\
   mHSE+V$_{\textup{w}}$	& 12.5 & 26.4 \\ \hline \hline
  V$_{\textup{Si}}$ in 4H-SiC &  $A_{\parallel }$ [MHz] & $A_{\perp}$ [MHz]  \\ \hline
    HSE06	& 246.9 & 32.4 \\
  HSE06+V$_{\textup{w}}$ &  233.1 &  32.8 \\
   Exp. \citep{Baur97} &  235.9 &  - \\
 \end{tabular}
\end{ruledtabular}
\end{table}

\subsection{V$_{\textup{Si}}$ in 4H-SiC}

The case of V impurity in 4H-SiC has been examined in our previous article \citep{Ivady13}, however, here, we reconsider this case with a more faithful treatment of the spurious electrostatic interaction of the charged point defect and carefully investigate the differences of the results of our scheme and the HSE06 functional. Additionally, we calculate the matrix elements of the hyperfine interaction of the vanadium and correlate the KS energy differences with excitation energies. 

The most favorable configuration of the vanadium impurity in 4H-SiC is as a substitutional defect at the silicon site. In the hexagonal 4H-SiC there are two different possible sites of a simple point defects, like V$_{\textup{Si}}$, known as $h$ and $k$ \citep{Ivady}. The electronic structures of these sites are approximately the same, therefore we only consider the $h$ site in the following.

\begin{figure}
\includegraphics[width=0.7\columnwidth]{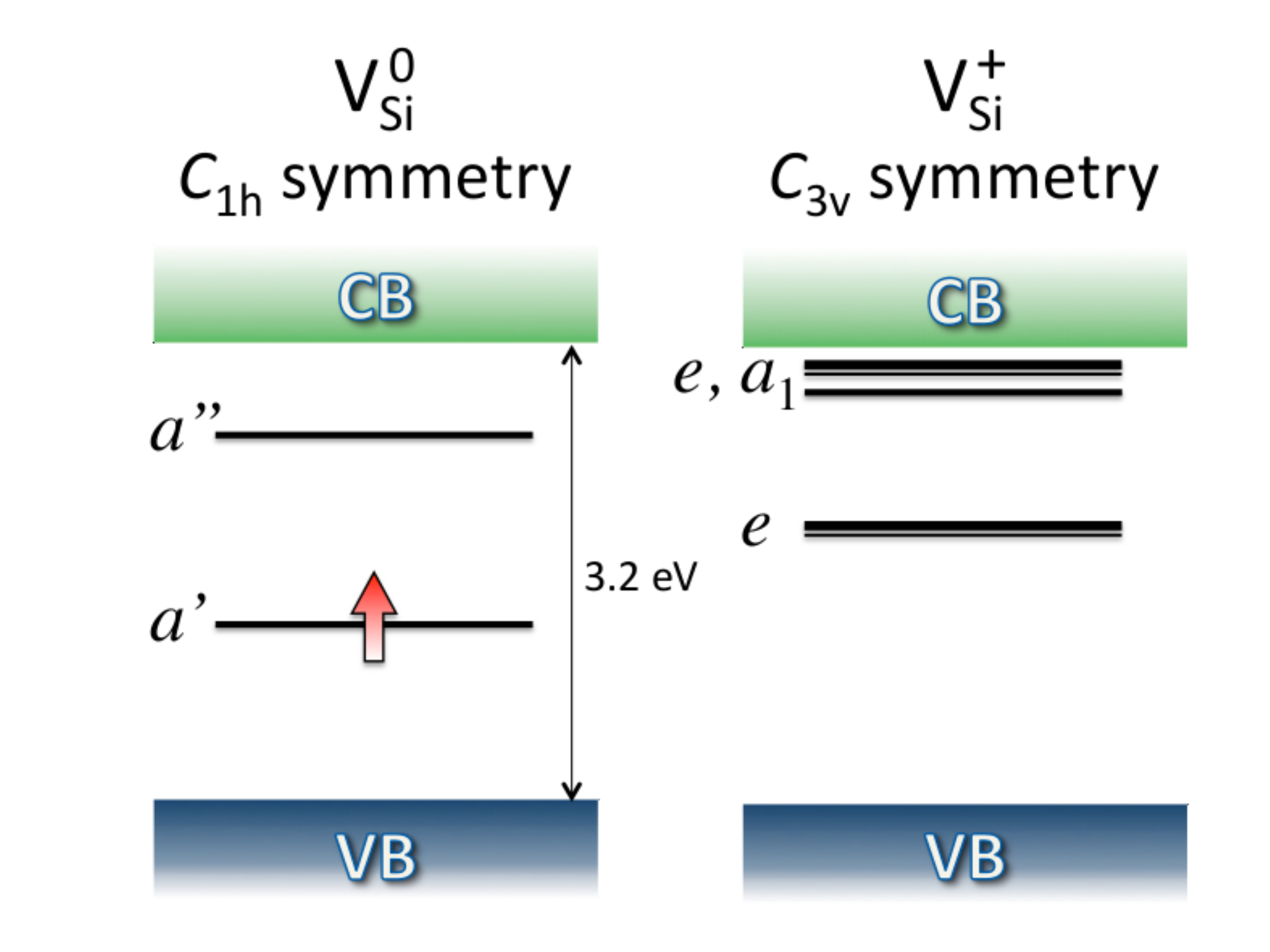}
\caption{\label{fig:V_lev}(Color online)  Schematic diagram of the defect orbitals of the neutral and positively charged V$_{\textup{Si}}$ point defect in 4H-SiC. }
\end{figure}

\begin{figure}
\includegraphics[width=0.8\columnwidth]{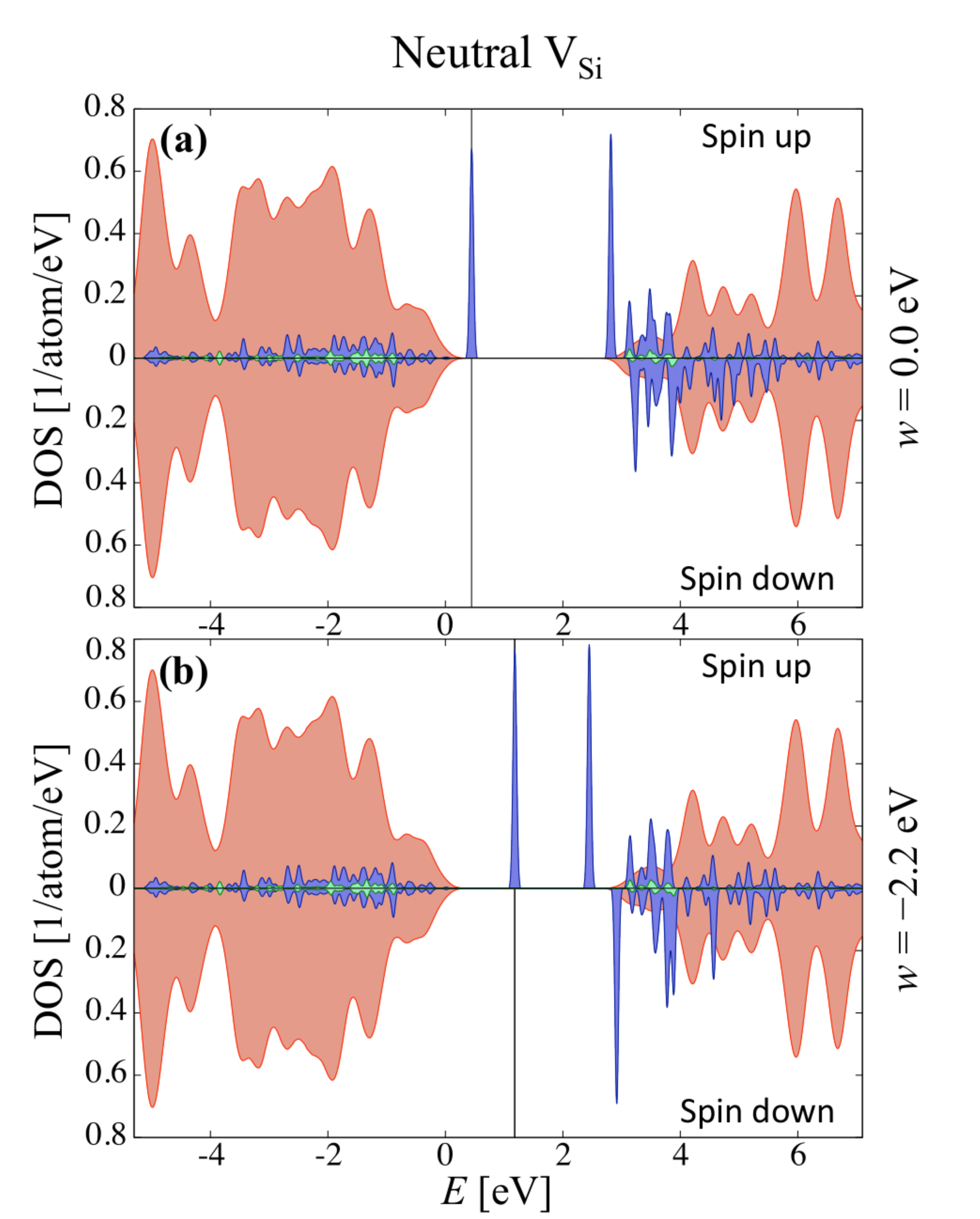}
\caption{\label{fig:pdos_V_0}  (Color online) Total and partial density of the states of the host 4H-SiC and the V impurity in the neutral V$_{\textup{Si}}$ point defect, respectively. The red filled curves show the total DOS of the host, while blue and green filled curves show the $d$ and $sp$ partial DOS of the vanadium.  These later curves were scaled up to be visible. Figure (a) and (b) show the results of the calculations with HSE06 and HSE06+V$_{\textup{w}}$ exchange correlation functional (see text for more explanation).}
\end{figure} 

Here, the previously discussed tight binding picture of the atomic orbitals can be adopted with the difference that the vacancy related states originally occupied by four electrons and to get fully occupied they capture four more electrons from the vanadium impurity and force it into a quasi V$^{\textup{4+}}$ configuration. Thus, in the neutral state of the V$_{\textup{Si}}$ defect, the atomic $d$-like orbitals are occupied by only one electron as shown in Fig.~\ref{fig:V_lev}. In the neutral charge state only the split lower lying $e$ state appears in the band gap of $3.1$~eV (see  Figs.~\ref{fig:V_lev}-\ref{fig:pdos_V_0}).

\begin{figure}
\includegraphics[width=0.85\columnwidth]{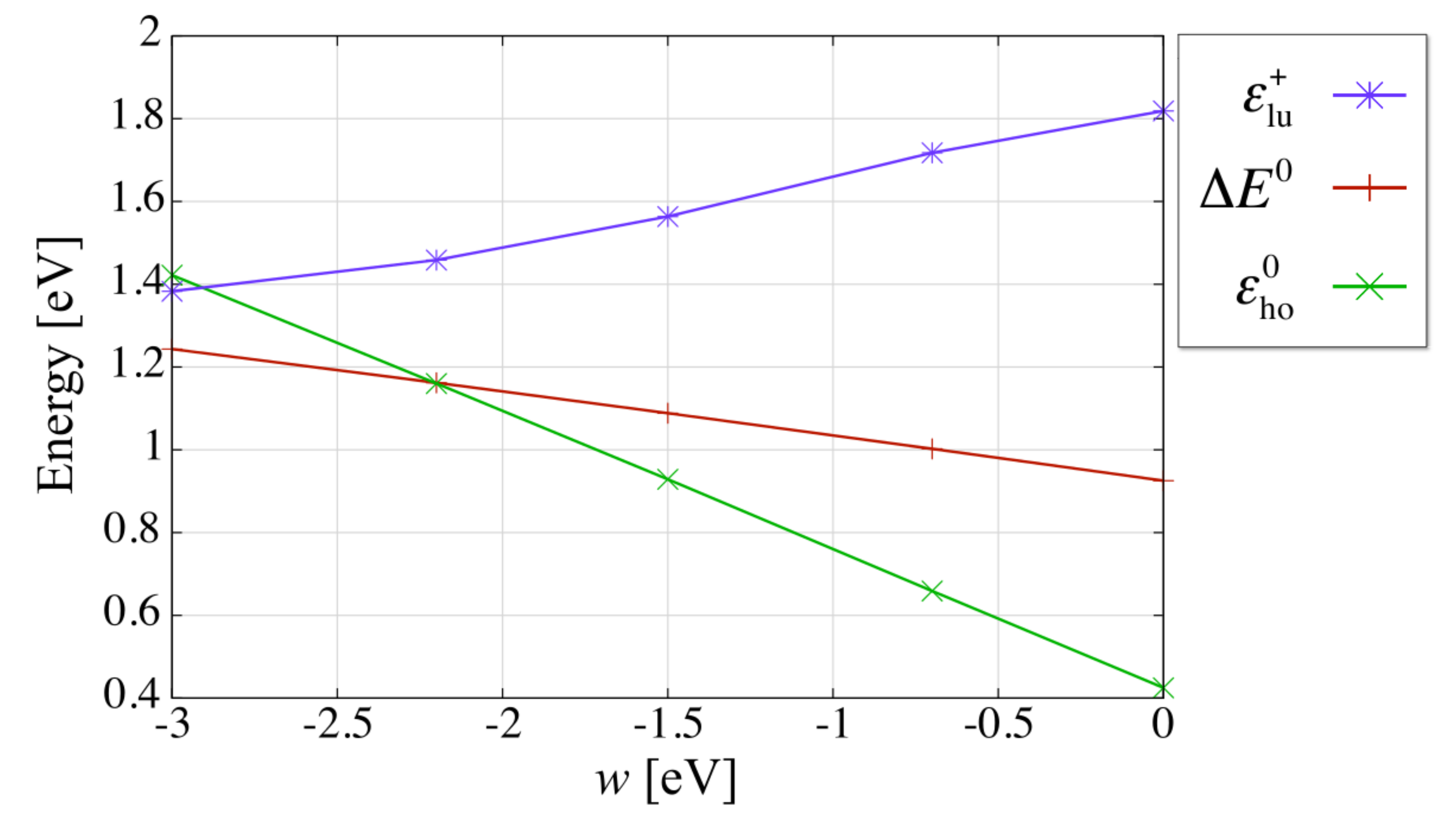}
\caption{\label{fig:V_NK} Variation of Kohn-Sham (KS) eigenvalues and the charge corrected total energy difference with respect to the strength of the correction parameter $w$ of HSE06+V$_{\textup{w}}$ method in the case V$_{\textup{Si}}$ in 4H-SiC (see text for more explanation). The variance of the highest occupied KS orbital in the neutral charge states and the lowest unoccupied state in the positively charged state are shown. The total energy difference is calculated from the total energies of the two charge states with applied charge correction. The valence band edge is chosen to possess the zero value on the energy scale. }
\end{figure}

To examine the accuracy of the HSE06 functional, we calculate the non-Koopmans' energy and its variation with respect to the strength of the correction potential $w$ (see Fig.~\ref{fig:V_NK}).  In the evaluation of Eq.~(\ref{eq:NK}) we use $\delta E^{\textup{cc}}_{+} = 0.11$~eV charge correction of the total energy of the positively charged supercell. The finite non-Koopmans' energy, $ E_{\textup{NK}} = 0.5$~eV, can be eliminated with the correction of $w = -2.2$~eV. As one may notice, the total energy does not change as rapidly as the KS eigenvalues as $w$ increases and the unoccupied state increase in energy, in contrast to the case of Cr$_{\textup{Al}}$ in w-AlN. This suggests that the contribution of the $d$-like orbitals to the valence band is less overestimated. One can also see that the charge correction of the KS eigenvalue of the unoccupied defect orbital is needed in the positive charge state to fullfil Eq.~(\ref{eq:me_3}), $\delta \varepsilon^{\textup{cc}}_{+} = -0.30$~eV.

To quantify the effect of the correction we compared the electronic structure of V$_{\textup{Si}}$ point defect as obtained with HSE06  and HSE06+V$_{\textup{w}}$ (Fig.~\ref{fig:pdos_V_0}). Due to the additional potential term the total energy difference is shifted upward with $0.24$~eV. The KS eigenvalues of the highest occupied and lowest unoccupied defect orbitals are increased with $0.74$~eV and decreased with $0.37$~eV in the neutral charge state, respectively. As a consequence the split of the $e$ state reduced from $2.369$~eV to $1.27$~eV. 

Differences of KS eigenvalues may not directly reflect the excitation energies, however, here we make an attempt to correlate the predictions of the obtained electronic structure with available photo luminescence (PL) measurements. The motivation for this comparison is that the non-empirical optimally tuned hybrids can reproduce excitation energies and quasiparticle spectra\citep{KronikPRL12,KronikJCTC12} and furthermore we could successfully correlate the KS eigenvalues of HSE06+V$_{\textup{w}}$ calculation with quasi particle energies\citep{Ivady13}. According to Magnusson \emph{et.al.}\citep{Magnusson00, Schneider90}, the ground state of the defect is located  $2.1\pm0.1$~eV below the conduction band edge and there is an inter impurity transition ($e \rightarrow e$) with $0.97$~eV energy in the case of V$_{\textup{Si}}$ defect at $h$ site. With the HSE06 functional one can predict $2.76$~eV and $2.369$~eV for the position of the highest occupied orbital and for the excitation energy. With the HSE06+V$_{\textup{w}}$ functional we obtained $2.0$~eV and $1.27$~eV for these quantities, which indicates remarkable improvement over the HSE06 results.  

The more careful treatment of the charge correction compared to our previous study reduces the refined $w$ parameter value with $0.5$~eV. Therefore, the calculated positive neutral charge transition level (+\textbar 0) is slightly shifted downwards with $0.06$~eV. However, this result is still improved compared with result of HSE06 calculation.  

For the observable densities, such as spin and total density, we have identified decreasing localization, but for the charge density of the highest occupied and the lowest unoccupied impurity states we again observed increased localization due to the applied correction V$_{\textup{w}}$. This may suggest that the overestimation of the linear combination of $d$-like impurity states and host related states is a common failure of hybrid functionals.

The calculated matrix elements of the hyperfine tensor are shown in Table~\ref{tab:hyperfine}. The values decrease in hybrid-DFT+V$\textup{w}$ which indicates delocalization. The comparison with the experimental value supports the need of the correction potential.

\section{Summary}

In summary, in this work we have revealed a formal connection for the treatment of localized states between two widespread first principles techniques, the hybrid-DFT and the DFT+$U$ method. The established connection allows us a formal motivation for the simultaneous combination of these two methods to overcome their limitations. This puts the hybrid-DFT+V$_{\textup{w}}$ method on formal footing as a technique to remedy the approximation of homogeneous and global screening of the Coulomb interaction introduced by the hybrid functionals, and makes it particularly suitable for simulations of systems with significantly different degree of localization of orbitals, like transition metal impurities in semiconductor host. In particular we have successfully demonstrated the advantages of this method in two different cases of Cr impurity in w-AlN and V impurity in 4H-SiC, where both quantitative and qualitative improvements were observed over the results of hybrid-DFT calculations.

\section{Acknowledgments} 

Discussion with P\'eter De\'ak are highly appreciated. Support from the Knut \& Alice Wallenberg Foundation ``Isotopic Control for Ultimate Materials Properties'', the Swedish Research Council (VR) Grants No.\ 621-2011-4426 and  621-2011-4249, the Swedish Foundation for Strategic Research  program SRL grant No.\ 10-0026, the Swedish National Infrastructure for Computing Grants No. SNIC 001/12-275 and No. SNIC 2013/1-331, and the ``Lend\"ulet program" of Hungarian Academy of Sciences is acknowledged. Use of the Center for Nanoscale Materials was supported by the U. S. Department of Energy, Office of Science, Office of Basic Energy Sciences, under Contract No. DE-AC02-06CH11357. R.A. acknowledges  support from the Linnaeus Environment at Link\"oping  on Nanoscale Functional Materials (LiLi-NFM) funded by VR.


%

\end{document}